\documentclass[twocolumn]{aastex63}
\pdfoutput=1 
\usepackage{natbib}
\usepackage[maxfloats=40]{morefloats}
\usepackage[T1]{fontenc}
\usepackage{color}
\usepackage{catoptions}
\usepackage{booktabs}
\usepackage{amsmath}
\usepackage[all]{hypcap}
\newcommand{\Rom}[1]{\uppercase\expandafter{\romannumeral #1\relax}}

\received{2020}
\revised{2020}
\accepted{}

\shorttitle{Diverse magnetic fields in B213 cores}
\shortauthors{Eswaraiah et al.}

\begin{document}
\makeatletter

\AuthorCallLimit=500

\title{Revealing the diverse magnetic field morphologies in Taurus dense cores with sensitive sub-millimeter polarimetry}

\correspondingauthor{Eswaraiah Chakali, FAST Fellow}
\email{eswaraiahc@nao.cas.cn,eswaraiahc@outlook.com}

\author[0000-0003-4761-6139]{Chakali Eswaraiah}
\affiliation{National Astronomical Observatories, Chinese Academy of Sciences, A20 Datun Road, Chaoyang District, Beijing 100012, People's Republic of China}

\author[0000-0003-3010-7661]{Di Li}
\affiliation{National Astronomical Observatories, Chinese Academy of Sciences, A20 Datun Road, Chaoyang District, Beijing 100012, People's Republic of China}
\affiliation{University of Chinese Academy of Sciences, Beijing 100049, People’s Republic of China}
\affiliation{NAOC-UKZN Computational Astrophysics Centre, University of KwaZulu-Natal, Durban 4000, South Africa}

\author[0000-0003-0646-8782]{Ray S. Furuya}
\affiliation{Tokushima University, Minami Jousanajima-machi 1-1, Tokushima 770-8502, Japan}
\affiliation{Institute of Liberal Arts and Sciences, Tokushima University, Minami Jousanajima-machi 1-1, 
Tokushima 770-8502, Japan}
 
\author[0000-0003-1853-0184]{Tetsuo Hasegawa}
\affiliation{National Astronomical Observatory of Japan, National Institutes of Natural Sciences, Osawa, Mitaka, Tokyo 181-8588, Japan}

\author[0000-0003-1140-2761]{Derek Ward-Thompson}
\affiliation{Jeremiah Horrocks Institute, University of Central Lancashire, Preston PR1 2HE, United Kingdom}

\author[0000-0002-5093-5088]{Keping Qiu} 
\affiliation{School of Astronomy and Space Science, Nanjing University, 163 Xianlin Avenue, Nanjing 210023, China}
\affiliation{Key Laboratory of Modern Astronomy and Astrophysics (Nanjing University), Ministry of Education, Nanjing 210023, China}
 
\author[0000-0003-0998-5064]{Nagayoshi Ohashi}
\affiliation{Academia Sinica Institute of Astronomy and Astrophysics, No.1, Sec. 4., Roosevelt Road, Taipei 10617, Taiwan}

\author[0000-0002-8557-3582]{Kate Pattle}
\affiliation{Centre for Astronomy, School of Physics, National University of Ireland Galway, University Road, Galway, Ireland H91TK33}

\author{Sarah Sadavoy}
\affiliation{Department for Physics, Engineering Physics and Astrophysics, Queen's University, Kingston, ON, K7L 3N6, Canada}

\author{Charles L. H. Hull}
\affiliation{National Astronomical Observatory of Japan, NAOJ Chile, Alonso de C\'{o}rdova 3788, Office 61B, 7630422, Vitacura, Santiago, Chile}
\affiliation{Joint ALMA Observatory, Alonso de C\'{o}rdova 3107, Vitacura, Santiago, Chile}
\affiliation{NAOJ Fellow}

% ...

\begin{abstract}

We have obtained sensitive dust continuum polarization observations at 850\,$\mu$m in the B213 region of Taurus using POL-2 on SCUBA-2 at the James Clerk Maxwell Telescope (JCMT), as part of the BISTRO (B-fields in STar-forming Region Observations) survey. These observations allow us to probe magnetic field (B-field) at high spatial resolution ($\sim$2000 au or $\sim$0.01 pc at 140 pc) in two protostellar cores (K04166 and K04169) and one prestellar core (Miz-8b) that lie within the B213 filament. Using the Davis-Chandrasekhar-Fermi method, we estimate the B-field strengths in K04166, K04169, and Miz-8b to be 38$\pm$14 $\mu$G, 44$\pm$16 $\mu$G, and 12$\pm$5 $\mu$G, respectively.
These cores show distinct mean B-field orientations. 
B-field in K04166 is well ordered and aligned parallel to the orientations of the core minor axis, outflows, core rotation axis, and large-scale uniform B-field, in accordance with magnetically regulated star formation via ambipolar diffusion taking place in K04166. B-field in K04169 is found to be ordered but oriented nearly perpendicular to the core minor axis and large-scale B-field, and not well-correlated with other axes.
In contrast, Miz-8b exhibits disordered B-field which show no preferred alignment with the core minor axis or large-scale field. We found that only one core, K04166, retains a memory of the large-scale uniform B-field.  The other two cores, K04169 and Miz-8b, are decoupled from the large-scale field. 
Such a complex B-field configuration could be caused by gas inflow onto the filament, even in the presence of a substantial magnetic flux.
	
\end{abstract}

\keywords{ISM: clouds -- ISM: individual objects (B213) -- ISM: magnetic fields -- polarization -- techniques: polarimetric stars: formation}

\section{Introduction}\label{sec:introd}

According to the filamentary paradigm of star-formation, low-mass stars predominantly form in dense
cores which are distributed in a chain-like fashion along  gravitationally unstable filamentary clouds 
\citep{Hartmann2002,TafallaHacar2015,Marshetal2016,Andreetal2014}. 
Magnetic field (B-field) is important at all scales during this process 
\citep{Shuetal1987,McKeeOstriker2007,Crutcher2012,Ward-Thompsonetal2020}. 
Nevertheless, the interplay between B-field, gravity, and turbulence in 
the formation of cores and their collapse to form stars is still a subject of investigation.

Studies of B-field on cloud scales with {\it Planck} 850 $\mu$m low-resolution ($\sim$5 arcmin or $\sim$0.2 at 140~pc) polarization observations and optical and near-infrared (NIR) polarimetry of background stars have revealed that low-density gas striations are 
mostly aligned to B-field and high-density filamentary structures are oriented perpendicular to B-field \citep{Chapmanetal2011,Alvesetal2008,Sugitanietal2010,PlanckCollaborationetal2016,WangJWetal2020}. These observations imply that material can accumulate along field lines and aid in the assembly of dense structures perpendicular to the B-field as a result of gravitational collapse and/or converging flows \citep[see][]{Ballesteros-Paredesetal1999,Hartmannetal2001,Soleretal2017}.

If the large-scale, uniform B-field is inherited down to core scale ($<$ 0.1 pc), they govern not only 
the contraction, stability, and collapse of the core \citep{MestelSpitzer1956,MouschoviasSpitzer1976}, 
but also the properties of circumstellar disk by help removing angular momentum via magnetic braking 
\citep{Mouschovias1991,Allenetal2003,LiZYetal2014}. According to the theory of isolated, low-mass star-formation via ambipolar diffusion
\citep{Mouschovias1991,Mouschoviasetal2006}, the gravitational collapse of a dense core is regulated by strong,
ordered B-field such that the core preferentially contracts along field lines. 
As a result, the core acquires an oblate-like structure over 10\,000 au scales. After gaining sufficient mass via B-field-mediated contraction,
initially the subcritical core becomes supercritical and eventually collapses under its own gravity. 
At this stage, the flux-freezing condition will no longer be valid due to efficient neutral-ion decoupling. 
As a result of this ambipolar diffusion, the B-field will acquire an hour-glass morphology 
on protostellar envelope scales, $<$ 1000 au \citep[e.g., ][]{GalliShu1993,Girartetal2006,Stephensetal2013}.
This model predicts a positive correlation between angle of the mean B-field and that of the 
minor axes of the filament and core, and the axes of both  pseudodisk symmetry and of the bipolar outflow 
\citep{FiedlerMouschovias1992,FiedlerMouschovias1993,GalliShu1993,Moczetal2017,HullZhang2019}.

Evidence for magnetically regulated star-formation through observations of coherent B-field across 
orders of magnitude in size scale (e.g., \citealt{LiHBetal2006}, \citealt{LiHBetal2009}, \citealt{Hulletal2014}) 
is not always the norm. 
Departure from coherency, especially at smaller scales, can occur in regions dominated by turbulence \citep[e.g.,][]{Hulletal2017a}, shocks from outflows \citep[e.g.,][]{Hulletal2017b}, gravity-driven gas flows \citep[e.g.,][]{Pillaietal2020}, stellar feedback driven by expanding ionization fronts from H\,{\sc ii} regions \citep{Arthuretal2011,Pattleetal2018,Eswaraiahetal2020}, or gas dynamics arise from gravitational collapse \citep{Chingetal2017,Chingetal2018}. 
These observations suggest that the very local environment can determine the morphology and role of B-field.

We emphasize here that B-field observations of low-mass dense cores (i) formed out of a single natal filament, (ii) characterized with ordered B-field at larger scales (sub-pc to several pc; see Figure 1(a)), (iii) having signpost of accretion flows \citep{Palmeirimetal2013,Shimajirietal2019}, and (iv) hosting pristine physical conditions, unaffected by any disruption by strong stellar feedback are sparse. The Taurus B213 is one of these rare regions, making the B213 cores the ideal laboratories to understand the role of B-field in the star formation process.

\begin{figure*}
\gridline{\fig{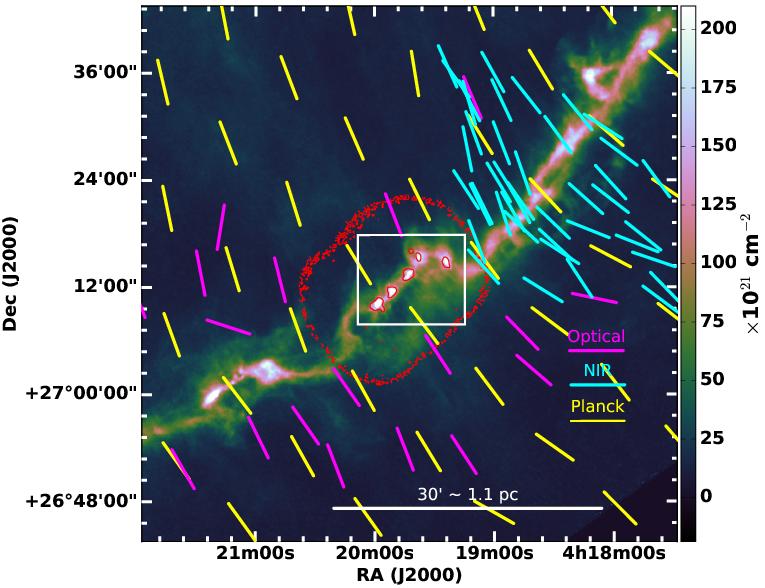}{0.5\textwidth}{(a)}
          \fig{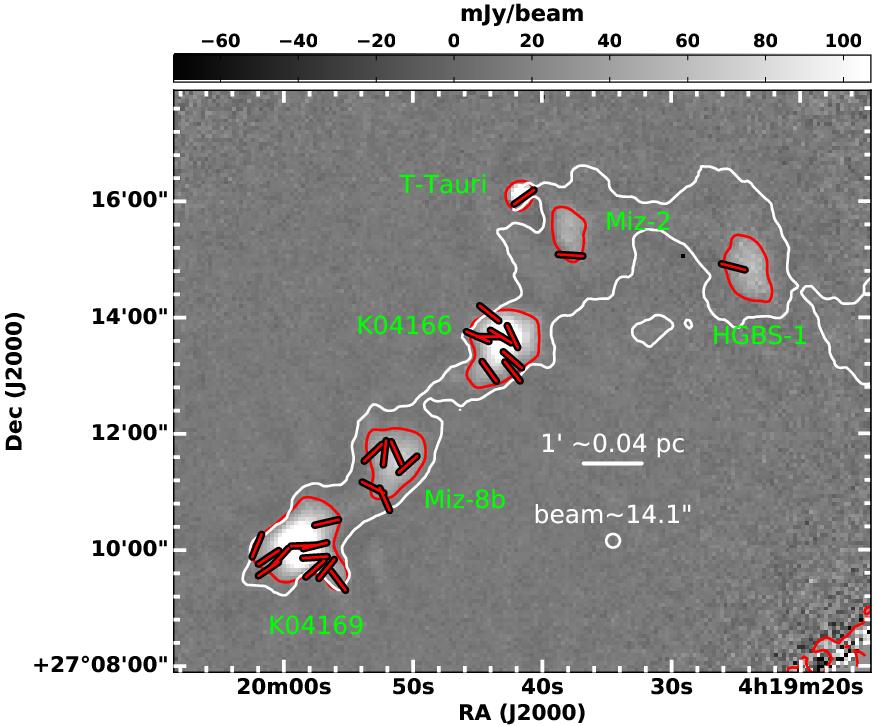}{0.5\textwidth}{(b)}}
\caption{(a) The overall structure of the B213 filament is shown using a high-resolution (18$\farcs$2) column density map taken from the {\it Herschel} Gould Belt Survey (HGBS) archive \citep{Palmeirimetal2013}. B213 is a part of the 10-pc-long large-scale filament LDN\,1495 in the Taurus molecular cloud. 
Measurements of the large-scale B-field orientation are overlaid: optical polarimetry measurements are shown as magenta segments \citep{Heyeretal1987,Heiles2000,Goodmanetal1990}; NIR polarimetry measurements, as cyan segments \citep{Goodmanetal1992,Chapmanetal2011}; and {\it Planck} 850 $\mu$m dust emission polarimetry measurements, as yellow segments. (b) Our inferred core-scale B-field geometry (red segments) superimposed on the area of our total intensity (Stokes I) map which contains the fragmented cores of B213. The extent of the map is shown as a white box in panel (a). Plotted segments correspond to a detection at a minimum of 3$\sigma$ in polarization fraction and 10$\sigma$ in total-intensity. Note that all the segments are shown with equal lengths to better display the B-field morphology. The white contour marks a column density of 1$\times10^{22}$ cm$^{-2}$ \citep{Palmeirimetal2013}, as measured in the {\it Herschel} data, which outlines the structure of the parental B213 filament, which has fragmented into the cores shown. Red contours drawn in both panels mark 10$\sigma$ total intensity (Stokes $I$) values, where 1$\sigma$~$=$~1.3 mJy beam$^{-1}$ is the rms noise. Note that the apparent $>$10$\sigma$ intensity values seen in low-column-density regions in panel (a) result from low exposure times at the edges of the POL-2 map, and so delineate the extent of our mapped area. Reference scale and beam size ($\sim$14$\farcs$1) are shown.}\label{fig:b213coldmap}
\end{figure*}

We conduct sensitive dust polarization observations at 850 $\mu$m towards B213, as part of the B-fields In STar-forming Region Observations (BISTRO; \citealt{Ward-Thompsonetal2017}) Survey, to resolve its B-field. BISTRO is a large program on 15-m James Clerk Maxwell Telescope (JCMT), making use of its SCUBA-2 camera and POL-2 polarimeter. B213 is a nearby (distance $\sim$140 pc; \citealt{Elias1978}) well-studied filament, which is part of the $\sim$10 pc filament LDN\,1495, as shown in Figure \ref{fig:b213coldmap}(a). B213 is fragmented into a chain of cores which are in the early evolutionary stages of low-mass star-formation (Figure \ref{fig:b213coldmap}(b)). These include  three prestellar cores, namely, Miz-8b, Miz-2, and HGBS-1 \citep{Mizunoetal1994,Marshetal2016}; two Class 0/I protostellar cores, IRAS 04166+2706 and IRAS 04169+2702 (\citealt{Ohashietal1997,Takakuwaetal2018,Tafallaetal2010}); and one evolved object, J04194148+2716070, classified as a Class II T Tauri star \citep{Davisetal2010}. We hereafter refer to IRAS 04166+2706 and IRAS 04169+2702 as K01466 and K04169, respectively \citep[cf.,][]{Kenyonetal1990,Kenyonetal1993}, adopting the core nomenclature of \citet{Braccoetal2017}.

Here, for the first time, we resolve the B-field in the three cores of B213 on 0.01~--~0.1 pc spatial scales. In this letter our key aims are to examine whether (i) B-field at scales $<$ 0.1~pc are coherent
with, or decoupled from, the uniform large-scale B-field, and (ii) the paradigm of magnetically regulated, isolated low-mass star-formation holds in these cores. This paper is organized as follows: Section \ref{sec:obs_data_redu} describes the observations and data reduction. Sections \ref{sec:results} and \ref{sec:discuss} present the results and discussion, respectively; and Section 5 summarizes our main findings.

\section{Observations and data reduction}\label{sec:obs_data_redu}

POL-2 observations of two fields in B213 were carried out as part of the JCMT BISTRO Survey (JCMT project code M16AL004) between 2017 November 05 and 2019 January 08. The two fields, shown in Figure  \ref{fig:pimap}, have a center-to-center angular separation of $\sim$5$^\prime$. The fields were each observed 20 times using the POL-2 DAISY mapping mode \citep{Hollandetal2013,Fribergetal2016}. This mode results in maps with a 12-arcminute diameter, of which the central $\sim$7 arcmin represents usable coverage, and so these two pointings represent a tightly-spaced mosaic. The observations were made in JCMT weather bands 1 and 2, with 225 GHz atmospheric opacity ($\tau_{225}$) varying between 0.02 and 0.06. The total exposure time for the two fields is $\sim$28 hrs (14 hrs in each of the two overlapping fields), resulting in one of the deepest observations yet made by the BISTRO Survey. 

The 850 $\mu$m POL-2 data were reduced using the pol2map routine recently added to SMURF \citep{Berryetal2005,Chapinetal2013}\footnote{ http://starlink.eao.hawaii.edu/docs/sun258.htx/sun258ss73.html}. The final mosaiced maps, calibrated in mJy beam$^{-1}$, are produced from co-added Stokes $I$, $Q$, and $U$ maps with pixel size of 4$\arcsec$, while the final debiased polarization vector 
catalog is binned to 12$\arcsec$ to achieve better sensitivity. 
The rms noise values in our Stokes 
$I$, $Q$, and $U$, and $PI$ maps, binned to a pixel size of 12$\arcsec$, are $\sim$1.3, $\sim$0.9, $\sim$0.9, and $\sim$1.0  mJy beam$^{-1}$, respectively.  $PI$ represents polarized intensity of the dust emission, debiased using the asymptotic estimator method; our $PI$ map is shown in Figure \ref{fig:pimap}. The instrumental polarization (IP) of POL-2 was corrected for using the `August 2019' IP model \citep{Fribergetal2018}. The POL-2 data reduction process is described in detail by \citet{Doietal2020} and \citet{Pattleetal2020}.


\section{Results}\label{sec:results}

\subsection{B-field on small scales}
\label{subsec:smallerBfields}

We present the data of 28 polarization measurements satisfying the criteria: (i) the ratio of intensity to its uncertainty $I/\sigma_{I} >$~10 and (ii) degree of polarization to its uncertainty $P/\sigma_{P} >$~3, where $P$~$=$~$PI$/$I$.  These measurements are listed in Table \ref{tab:poldata}. The resulting $PI$ 
within the core boundaries (see Appendix \ref{subsec:weakpi}) ranges from $\sim$2 to $\sim$4 mJy beam$^{-1}$ with a median uncertainty in polarized intensity, $\sigma_{PI}$, of 0.64 mJy beam$^{-1}$. The polarization fraction ranges from $\sim$0.8 to $\sim$18\% with a median value of $\sim$7\%. The B213 cores are characterized by weak dust emission ($\sim$12~--~318 mJy beam$^{-1}$) as well as weak polarized emission in comparison to the other regions studied regions by the BISTRO program \citep{Ward-Thompsonetal2017,Kwonetal2018,Soametal2018,WangJWetal2019,Coudeetal2019,LiuJunhaoetal2019,Pattleetal2019,Doietal2020}. 

Assuming a distance to Taurus of 140 pc, our observations allow us to delineate B-field in B213 on scales ranging from $\sim$2000 au ($\sim$0.01 pc) up to $\sim$0.25 pc, the length over which the cores K04166, Miz-8b and K04169 are distributed.
The resulting B-field geometry, based on the 28 polarization measurements (cf., Table \ref{tab:poldata}), is shown in Figure \ref{fig:b213coldmap}(b). Since the three cores, T-Tauri, Miz-2, and HGBS-1, have only a single measurement 
each (and also because the background noise dominates at the locations of these cores -- cf., Appendix \ref{subsec:weakpi}), 
we exclude them from further analysis and discussion. The overall B-field morphology appears to be uniform within K04166 and K04169, but the mean field directions are offset by $\sim$90$\degr$ from one another. In contrast, the B-field morphology in Miz-8b is complex. 

\begin{figure*}
\centering
\resizebox{18cm}{16cm}{\includegraphics{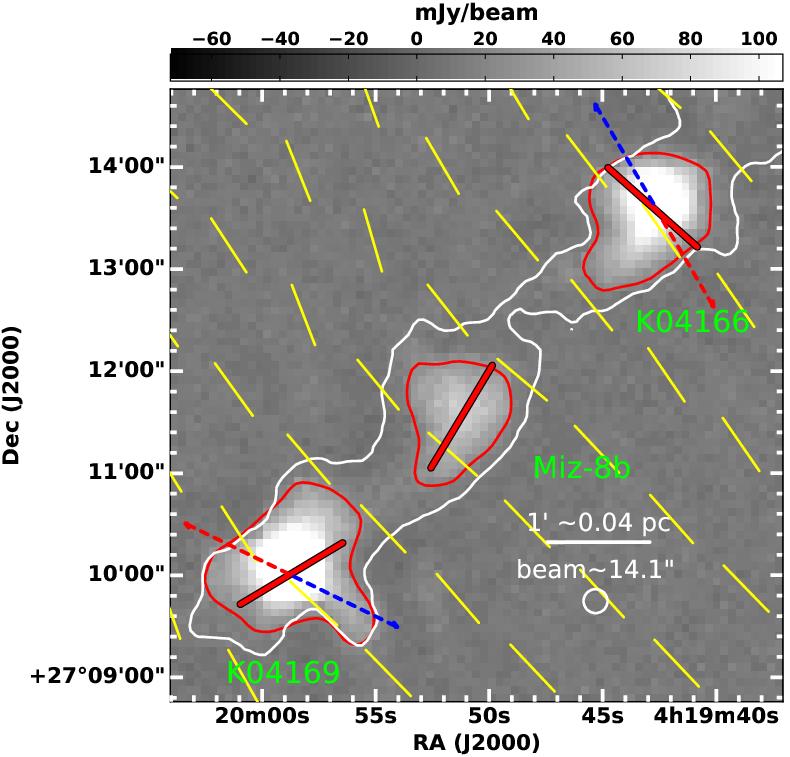}}
        \caption{Same as Figure \ref{fig:b213coldmap}(b) but now we only show the mean B-field orientation in the three cores K04166, Miz-8b, and K04169, based on the weighted mean of the position angles which we measure. The blue and red dashed arrows denote the protostellar outflows (lengths are not to scale) emanated from K04166 and K04169. The red contour around each core is drawn at $I$~$=$~13 mJy beam$^{-1}$, corresponding to 10$\sigma$ in total intensity. The large-scale B-field morphology, as determined from the over-sampled {\it Planck} 850 $\mu$m
        polarization data (pixel size 1$\arcmin$), is shown as yellow segments. The white contour is as described in Figure \ref{fig:b213coldmap}(b).\label{fig:meanBfieldmap}}
        \end{figure*}

We compute the weighted mean position angle
of the core B-field, $\bar{\theta}_{\mathrm{core, B}}$, using uncertainties in polarization angle as weights. These values are given in Table \ref{tab:paramscl12}. The $\bar{\theta}_{\mathrm{core, B}}$ along with the low-resolution B-field morphology based on 
{\it Planck}~850 $\mu$m polarization data, is shown in Figure \ref{fig:meanBfieldmap}. Table \ref{tab:paramscl12} lists the offset 
between $\bar{\theta}_{\mathrm{core, B}}$ and the large scale mean B-field orientation ($\theta_{\mathrm{B}}^{\mathrm{largescale}}$; see Appendix \ref{subsec:largescaleBfields}) based on multi-wavelength polarimetry. 
Also listed are the the offset between $\bar{\theta}_{\mathrm{core, B}}$ 
and the position angle of each core's major axis
($\theta_{\mathrm{core}}$; see Appendix \ref{appendixsec:geomass}).

Interestingly, we see completely different B-field geometry in each of the three cores. The B-field in K04166 lies roughly parallel to the large-scale field (or perpendicular to the filament), while that in K04169 lies roughly perpendicular to the large-scale field (or roughly parallel to the filament).
The field direction in Miz-8b lies roughly half-way between the other two, albeit with a larger standard deviation in B-field orientations (35$\degr$; see Table \ref{tab:paramscl12}). Hence, we see that the core-scale B-field, in a set of cores spanning $\sim6\arcmin$, or $\sim$0.25 pc, appear to be rather complex.

Furthermore, we observe a good alignment between core B-field ($\sim$48$\degr$) and outflows ($\sim$33$\degr$) in K04166 (Figures \ref{fig:meanBfieldmap} and \ref{fig:igbfield4446}(d)), consistent with studies by \citet{Davidsonetal2011,Chapmanetal2011}. In contrast, we see a misalignment between core mean B-field ($\sim$121$\degr$) and outflows ($\sim$58$\degr$) in K04169 (Figures \ref{fig:meanBfieldmap} and \ref{fig:igbfield4446}(e)), in accordance with studies by \citet{Hulletal2013,Hulletal2014,HullZhang2019,Yenetal2020}. 
Despite the fact that these two cores lie within $\sim$0.25 pc of each other, they exhibit different B-field/outflow orientations.


\subsection{B-field strength}\label{subsec:b_field_strength}  

Based on the assumption that turbulence-induced Alfv\'{e}n waves can distort B-field orientations, 
the plane-of-sky component of the B-field strength ($B_{pos}$) can be estimated using the 
Davis-Chandrasekhar-Fermi relation \citep[][DCF method]{Davis1951,ChandrasekharFermi1953}: 

\begin{equation}\label{eq:bfield_acf}
        B_{\mathrm{pos}} \simeq Q \sqrt{4\pi\,\mu m_{\mathrm{H}}\,n_{\mathrm{H_{2}}}}\,\left(\frac{\delta_{\mathrm{V_{NT}}}}{\delta_{\mathrm{\theta}}}\right), 
\end{equation}

where $n_{\mathrm{H_{2}}}$ is the gas number density, $\delta_{\mathrm{V_{NT}}}$ is the non-thermal gas velocity dispersion, and $\delta_{\theta}$ is the dispersion in polarization angles about the mean B-field orientation. 
$Q$ is a factor accounting for line-of-sight and beam dilution 
effects, which we take as 0.5 based on studies using synthetic polarization maps generated from numerically
simulated clouds \citep{Ostrikeretal2001}, which suggest that without this correction factor, DCF-measured B-field strength is overestimated by a factor of 2 when the angular dispersion in the B-field is $\leqslant$25$\degr$.

As illustrated in Appendix \ref{appendixsec:geomass}, we have used the 850-$\mu$m Stokes I map to extract core dimensions, column and number densities, and masses. To quantify the non-thermal velocity dispersion induced by the turbulence, we estimated the 
average velocity dispersion ($\delta_{\mathrm{V_{LSR}}}$) from archival 
N$_{2}$H$^{+}$ (1~--~0) data \citep{Punanovaetal2018}\footnote{Data can be found at \url{http://cdsarc.u-strasbg.fr/ftp/J/A+A/617/A27/fits/}}, obtained using  the IRAM 30-m telescope. The spatial and velocity resolutions of the N$_{2}$H$^{+}$ data are 
26$\farcs$5 and 0.063 km s$^{-1}$, respectively. 
The thermal contributions to the observed velocity dispersions ($\delta_{\mathrm{V_{T}}}$) are estimated (based on the mean dust temperatures of the cores given in Table \ref{tab:paramscl12}). These components are quadratically subtracted from the observed velocity dispersions 
($\delta_{\mathrm{V_{LSR}}}$) to obtain non-thermal velocity 
dispersions ($\delta_{\mathrm{V_{NT}}}$). 
The angular dispersion in the B-field is calculated using the relation for the inverse-variance-weighted standard 
deviation of B-field \citep[e.g,][]{WangJWetal2020}. 
These estimated parameters are listed in Table \ref{tab:paramscl12}.

Using equation \ref{eq:bfield_acf} and the parameters listed above, the
B-field strength is estimated to be 38$\pm$14 $\mu$G for K04166, 44$\pm$16 $\mu$G for K04169, and 12$\pm$5 $\mu$G for Miz-8b. Since the majority of the B-field segments in K04166 and K04169 are confined to the core radii $\sim$20~--~50$\arcsec$, the B-field strengths in these cores are mainly valid to the core-envelopes. Further, we caution here that the B-field strength of Miz-8b could be highly uncertain because of the limited number of B-field segments, and hence the larger angular dispersion, used in the DCF method. 
The current estimations are similar to 
the B-field strengths of $\sim$10~--~100 $\mu$G estimated in relatively unperturbed 
low-mass star-forming regions \citep{Crutcheretal2004,Chapmanetal2011,Crutcher2012}, 
and two orders of magnitude less than the $\sim$1 mG values estimated in massive star forming regions 
\citep[eg.,][]{CurranChrysostomou2007,Hildebrandetal2009,Pattleetal2017,LiuJunhaoetal2020}.

We can use our estimated B-field strength to infer 
the dynamic state and physical properties of the cores (see Table \ref{tab:paramscl12}). First, we estimate the magnetic and turbulent pressures using the 
relations $P_{B}$~$=$~$B^{2}/8\pi$ and $P_{\mathrm{turb}}$~$=$~$\rho{\delta_{\mathrm{NT}}}^{2}$, respectively. Second, we estimate the Alfv\'{e}nic Mach number using the relation 
$M_{A}$~$=$~$\sqrt{3} (\frac{\delta_{\mathrm{NT}}}{V_{\mathrm{A}}})$,
where Alfv\'{e}n velocity $V_{\mathrm{A}}$~$=$~$\frac{B_{\mathrm{los}}}{\sqrt{4\pi\,\rho}}$
(where $\rho$~$=$~$n_{H_{\mathrm{2}}}\,\mu\,m_{\mathrm{H}}$). Third, we use the mass-to-magnetic flux ratio to infer how important is the B-field in comparison to gravity. We measure the mass-to-flux ratio in units of the critical value, as described in Appendix \ref{subsec:magnetic_criticality}. Finally, we estimate the rotational energy of each core to determine how rotation may influence the B-field in Appendix \ref{appendix:mag2rot}. The derived energy values, along with all other parameters, are listed in Table \ref{tab:paramscl12}. 
 
\section{Discussion}\label{sec:discuss}

Since the two protostellar cores, K04166 and K04169, are at a similar evolutionary stage and share similar characteristics (see Table \ref{tab:paramscl12}), we discuss their energy parameters and gas kinematics with reference to the differences in B-field morphology in Section \ref{subsec:comparisonparametrs}. These aspects for the prestellar core Miz-8b are addressed in section \ref{subsubsec:miz8bsfr}.

\subsection{K04166 and K04169}\label{subsec:comparisonparametrs}

Magnetic-to-turbulent pressure ratio is seen to be $\sim$1 in both cores (see Table \ref{tab:paramscl12}). This suggests that B-field and turbulence are near equilibrium with each other. Equivalently, the Alfv\'{e}nic Mach number ($\sim$1) suggests that turbulent motions 
are trans-Alfv\'{e}nic. Therefore turbulent motions are not dominant over, and so do not shape the morphology of, the B-field in these cores. 
The mean mass-to-flux ratio criticality of the cores, $\lambda$, is found to be  $\sim$1, 
suggesting that the core-envelopes may be magnetically critical and marginally supported by B-field. The ratio of rotational to magnetic energy (cf., Appendix \ref{appendix:mag2rot}), 
$E_{\mathrm{rot}}/E_{\mathrm{mag}}$~$\ll$~1, which infers that the core rotational energy is too weak to alter the B-field orientation.

\begin{figure*}
\centering
	  \gridline{\fig{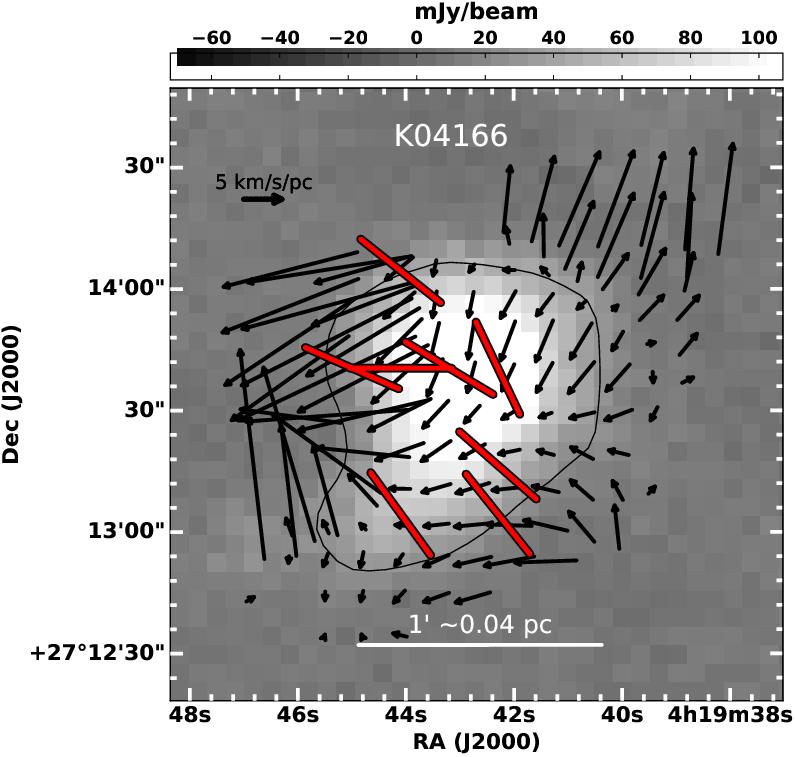}{0.35\textwidth}{(a)}
	  \fig{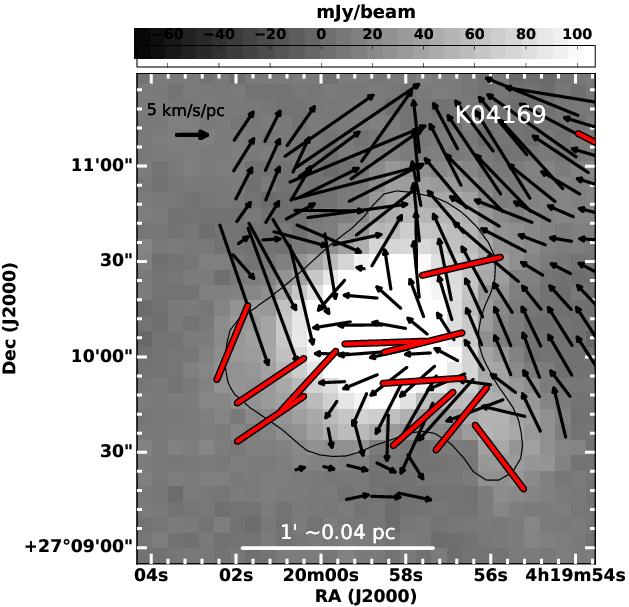}{0.35\textwidth}{(b)}
	  \fig{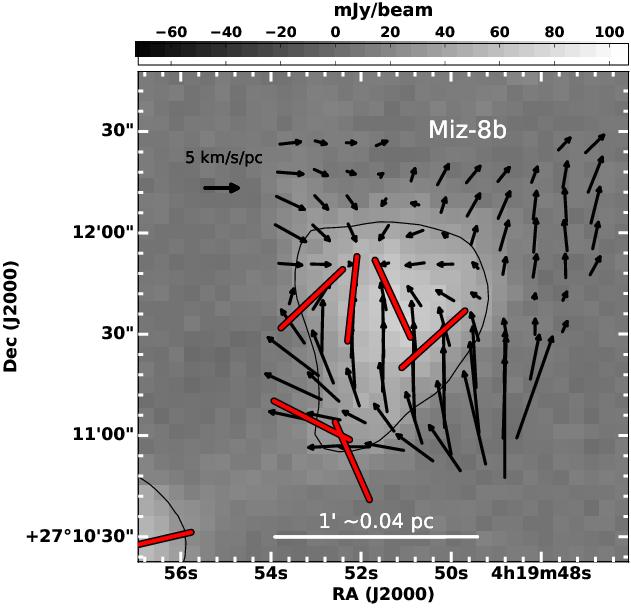}{0.35\textwidth}{(c)}}
	  \gridline{\fig{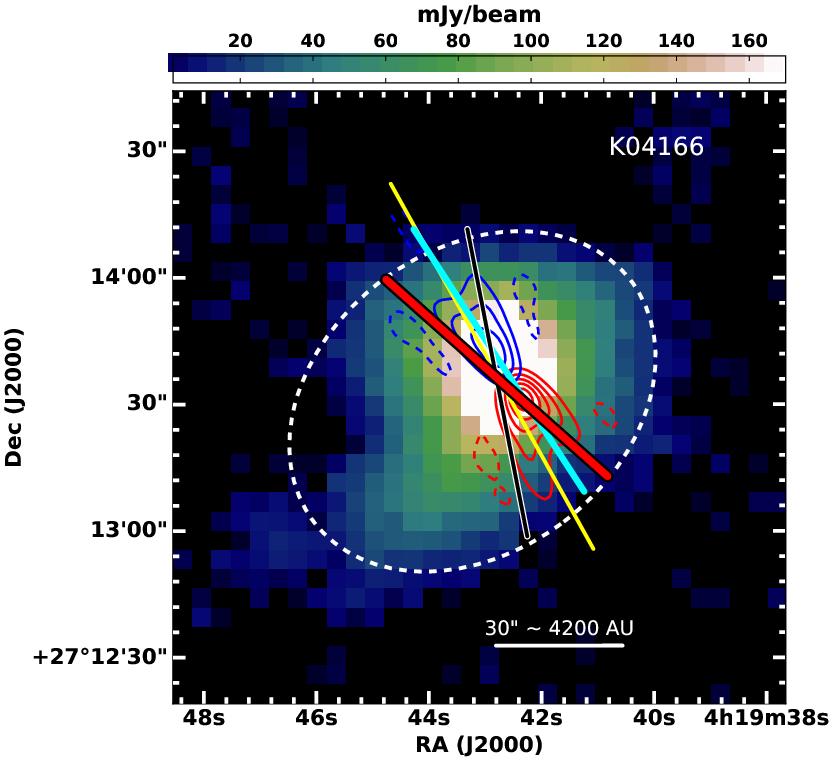}{0.35\textwidth}{(d)}
	  \fig{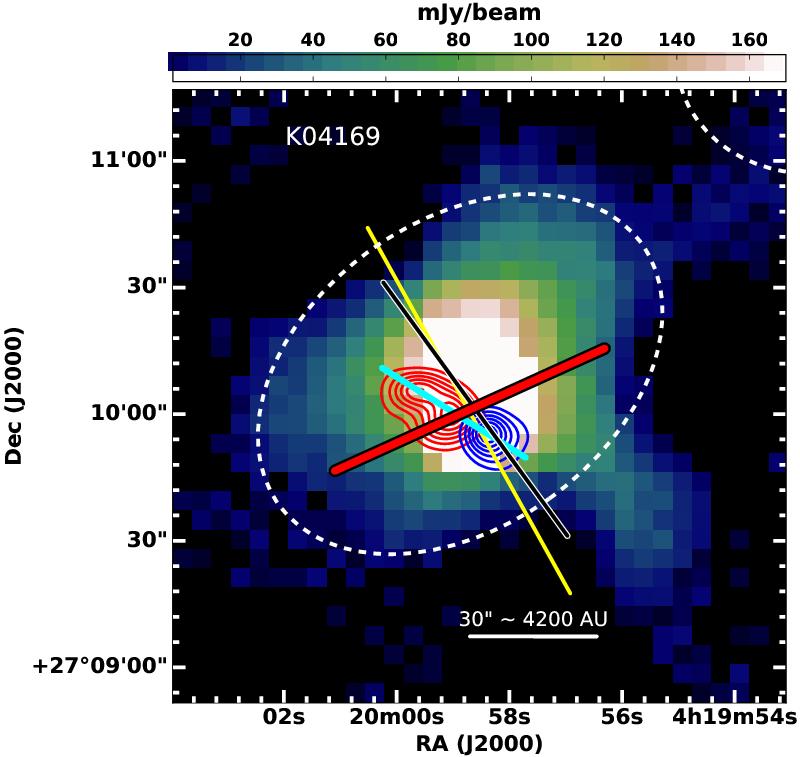}{0.35\textwidth}{(e)}
	 \fig{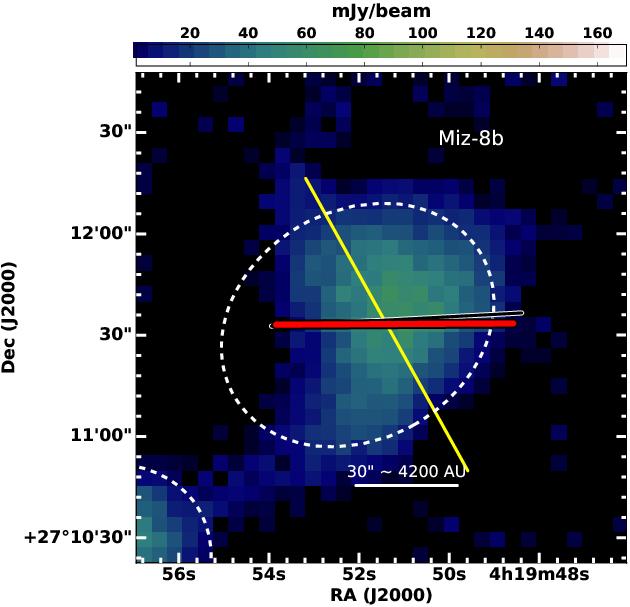}{0.35\textwidth}{(f)}}
	\caption{(Top panels) Velocity gradient (VG) maps for (a) K04166, (b) K04169, and (c) Miz-8b. The length and angle of each black arrow corresponds to the magnitude and direction, respectively, of the local VG. A reference arrow with the units of km s$^{-1}$ pc$^{-1}$ is shown on each plot. Red segments show the B-field orientations from Figure~\ref{fig:b213coldmap}(b). (Bottom panels) The weighted mean B-field geometry (thick red segments; same as Figure~\ref{fig:meanBfieldmap}) in the cores of (d) K04166, (e) K04169, and (f) Miz-8b overlaid on the Stokes I map (color scale). The large-scale {\it Planck} mean B-field direction is shown as a yellow segment on each core. Red and blue contours show the red-shifted and blue-shifted $^{12}$CO (2--1) emission tracing the bipolar outflow in the center of each core. These were obtained with the ALMA 7-m array by \citet{Tokudaetal2020}. The cyan line denotes the mean position angle of each outflow (33$\degr$ for K04166 and 58$\degr$ for K04169). Thin black line represents the position angle of the core rotation-axis. The integrated velocity ranges are $-$10 to 0 and 10 to 20 km s$^{-1}$ for the blue- and red-shifted outflows of K04166, and $-$5 to 4 and 11 to 17 km s$^{-1}$ for the blue- and red-shifted outflows of K04169. The lowest and subsequent contour steps are 0.4 and 1.2 K km s$^{-1}$ , respectively. The
        angular resolution of the ALMA data is $6.8\arcsec\times6.5\arcsec$.}
        \label{fig:igbfield4446}
        \end{figure*}

Our analysis indicates that there is an equipartition among magnetic, turbulent, and gravitational energies in the core-envelopes of K04166 and K04169. Then the question arises as to why the mean B-field orientation in the two cores are different from each other. We use the morphological correspondence between N$_{2}$H$^{+}$ velocity gradients and B-field, as shown in Figures \ref{fig:igbfield4446}(a) \& (b), to shed light on this.

The velocity field in K04166 as inferred from velocity gradient map is well defined, fairly uniform, and is almost perpendicular to the B-field segments. This could be interpreted as bulk core rotation, with the angular momentum (or core rotation-axis with PA $\sim$11$\degr$) being parallel to the B-field direction. In addition, the outflow is well-collimated and exhibits extremely high velocity (EHV) components \citep{WangLYetal2014}, suggesting a possible role of B-field in channeling the outflow and transporting energy and angular momentum away from the rotating circumstellar disk. The position angles of the core (and filament) minor axis ($\sim$37$\degr$), core rotation-axis ($\sim$11$\degr$), and bipolar outflows ($\sim$33$\degr$) are all roughly aligned with both  $\bar{\theta}_{\mathrm{core, B}}$ (48$\degr$) and $\theta_{\mathrm{B}}^{\mathrm{largescale}}$ (29$\degr$) to within $\sim$30$\degr$, as shown in Figure \ref{fig:igbfield4446}(d). This strong geometrical correspondence suggests that B-field, which is inherited from the large-scale uniform B-field, have played a significant role in core evolution by allowing gas contraction along field lines to form the core, subsequently governing its collapse via ambipolar diffusion, and finally collimating the outflows. These signatures are in accordance with paradigm of low-mass star-formation process driven by ambipolar diffusion in K04166. However, we could not trace an hour-glass morphology in the inner core (radii $<$ 20$\arcsec$ or $<$ 2800 au) due to limited resolution ($14\farcs1 \sim$ 2000 au).

On the other hand, the velocity gradient map in K04169 
appears to be rather complex, which displays two converging flows, from the northeast and the southwest (Figure \ref{fig:igbfield4446}(b)). 
Counter-rotation between the disk and the envelope in K04169 is also reported \citep{Takakuwaetal2018}. We see that core mean B-field ($\bar{\theta}_{\mathrm{core, B}}$ $\sim$121$\degr$) is nearly aligned parallel to the mean orientation of velocity gradient ($\theta_{\mathrm{G}}$~$ \sim$126$\degr$; cf., Table \ref{tab:paramscl12}). We suggest that this complex gas flows might have altered the B-field
from being parallel to the core minor axis in the earlier stage to current perpendicular configuration in K04169. 
This might have also caused the misalignment of outflows (PA $\sim$58$\degr$), core rotation-axis (PA $\sim$36$\degr$), and core minor axis (PA $\sim$36$\degr$) with $\bar{\theta}_{\mathrm{core, B}}$ as shown in Figure \ref{fig:igbfield4446}(e) (see Table \ref{tab:paramscl12} for more details). 

\subsection{Miz-8b}\label{subsubsec:miz8bsfr}
 
 Unlike those in K04166 and K04169, Miz-8b has a disordered B-field. 
 It has a magnetic-to-turbulent pressure ratio of $\sim$0.3 and a Alfv\'{e}nic Mach number of $\sim$2 (cf., Table \ref{tab:paramscl12}), which suggests that turbulence is super-Alfv\'{e}nic and dominates over B-field.
 Cores formed in a weakly magnetized, turbulent cloud  would have a chaotic B-field configuration, because of the dominance of turbulent eddies over structural dynamics and field lines \citep{Stoneetal1998,Ballesteros-Paredesetal1999b,MacLowKlessen2004,LiZYetal2014}. 
We suggest that B-field in Miz-8b are complex because of the dominance of turbulent flows 
(Figure \ref{fig:igbfield4446}(c)). As a result B-field is decoupled from large-scale ordered field (Figure \ref{fig:igbfield4446}(f)).
Since the inferred B-field strength in Miz-8b is weaker (12$\pm$5 $\mu$G), it will not support the core against gravity as the mass-to-flux ratio is found to be supercritical ($\lambda \sim$3).

\section{Summary}\label{sec:summary}

We have performed deep dust polarization observations towards the Taurus B213 filament at 850~$\mu$m 
using SCUBA-2 and POL-2 on the JCMT as part of the BISTRO survey.
We successfully detected polarized signal in, and studied in detail the B-field of, two proto-stellar cores (K04166 and K04169) and one prestellar core (Miz-8b) on scales from 2000~au to 0.25~pc.The main findings of this work are as follows:

\begin{enumerate}

\item Despite having (i) ordered B-field on large scales, (ii) quiescent physical conditions, and (iii) formed out of the same natal filament, the three B213 cores exhibit diverse magnetic field properties.

\item Among the three cores, only one, K04166, retains a memory of the large-scale B-field, with a field orientation consistent with those seen on larger scales.  The other two cores appear to have decoupled from the large-scale field.

\item Using the Davis-Chandrasekhar-Fermi method, we estimate the B-field strengths in K04166, K04169, and Miz-8b to be 38$\pm$14 $\mu$G, 44$\pm$16 $\mu$G, and 12$\pm$5 $\mu$G, respectively.  The associated magnetic energies are in equipartition with both turbulent and gravitational energies in the core envelopes of K04166 and K04169, while being much smaller than the turbulent energy in the core Miz-8b. 

\item Based on the correlation between the position angle of the core-scale B-field and the large-scale field, the minor axis of the core, outflows, and the core rotation axis, we suggest that the formation and evolution of K04166 are regulated by B-field, consistent with the paradigm of low-mass star-formation via ambipolar diffusion. However, as revealed by their complex velocity fields, the evolution of the other two cores, K04169 and Miz-8b, could be regulated by converging accretion flows and turbulent motions respectively.

\end{enumerate}

We suggest that cores formed in a magnetically regulated molecular cloud may not necessarily retain a memory of the large-scale B-field of the cloud in which they form. Instead, localized differences in gas kinematics, which probably arise due to gas inflows onto the filament, can affect the role of B-field in the star formation process, and the subsequent properties of the forming systems.

\facilities{JCMT}

\clearpage

\acknowledgments

{\scriptsize{
We thank the referee for constructive suggestions which have improved the content and flow of this paper. This work is supported by the National Natural Science Foundation of China (NSFC) grant Nos. 11988101, 11725313, and U1931117, and the International Partnership Program of Chinese Academy of Sciences grant No. 114A11KYSB20160008. This work is also supported by Special Funding for Advanced Users, budgeted and administrated by the Center for Astronomical Mega-Science, Chinese Academy of Sciences (CAMS). C.E. thanks Tokuda Kazuki for kindly providing us with the ALMA data on the protostellar outflows. C.E. thanks Sihan Jiao and Yuehui Ma for help with the analyses. K.Q. acknowledges the support from National Natural Science Foundation of China (NSFC) through grant U1731237. C.W.L. is supported by the Basic Science Research Program through the National Research Foundation of Korea (NRF) funded by the Ministry of Education, Science and Technology (NRF-2019R1A2C1010851). N.O. is supported by Ministry of Science and Technology (MOST) grant MOST 109-2112-M-001-051 in Taiwan. C.L.H.H. acknowledges the support of the NAOJ Fellowship and JSPS KAKENHI grants 18K13586 and 20K14527. W.K. was supported by the New Faculty Startup Fund from Seoul National University. D.J. is supported by the National Research Council of Canada and by a Natural Sciences and Engineering Research Council of Canada (NSERC) Discovery Grant. A.S. acknowledges support from the NSF through grant AST-1715876. 
The James Clerk Maxwell Telescope is operated by the East Asian Observatory on behalf of
The National Astronomical Observatory of Japan; Academia Sinica Institute of Astronomy and Astrophysics;
the Korea Astronomy and Space Science Institute; Center for Astronomical Mega-Science. 
Additional funding support is provided by the Science and Technology Facilities Council of the
United Kingdom and participating universities in the United Kingdom, Canada and Ireland.}}

\bibliographystyle{aasjournal}
\bibliography{myref}

\begin{thebibliography}{}
\expandafter\ifx\csname natexlab\endcsname\relax\def\natexlab#1{#1}\fi
\providecommand{\url}[1]{\href{#1}{#1}}
\providecommand{\dodoi}[1]{doi:~\href{http://doi.org/#1}{\nolinkurl{#1}}}
\providecommand{\doeprint}[1]{\href{http://ascl.net/#1}{\nolinkurl{http://ascl.net/#1}}}
\providecommand{\doarXiv}[1]{\href{https://arxiv.org/abs/#1}{\nolinkurl{https://arxiv.org/abs/#1}}}

\bibitem[{{Allen} {et~al.}(2003){Allen}, {Li}, \& {Shu}}]{Allenetal2003}
{Allen}, A., {Li}, Z.-Y., \& {Shu}, F.~H. 2003, \apj, 599, 363,
  \dodoi{10.1086/379243}

\bibitem[{{Alves} {et~al.}(2008){Alves}, {Franco}, \& {Girart}}]{Alvesetal2008}
{Alves}, F.~O., {Franco}, G.~A.~P., \& {Girart}, J.~M. 2008, \aap, 486, L13,
  \dodoi{10.1051/0004-6361:200810091}

\bibitem[{{Andr{\'e}} {et~al.}(2014){Andr{\'e}}, {Di Francesco},
  {Ward-Thompson}, {Inutsuka}, {Pudritz}, \& {Pineda}}]{Andreetal2014}
{Andr{\'e}}, P., {Di Francesco}, J., {Ward-Thompson}, D., {et~al.} 2014, in
  Protostars and Planets VI, ed. H.~{Beuther}, R.~S. {Klessen}, C.~P.
  {Dullemond}, \& T.~{Henning}, 27,
  \dodoi{10.2458/azu_uapress_9780816531240-ch002}

\bibitem[{{Arthur} {et~al.}(2011){Arthur}, {Henney}, {Mellema}, {de Colle}, \&
  {V{\'a}zquez-Semadeni}}]{Arthuretal2011}
{Arthur}, S.~J., {Henney}, W.~J., {Mellema}, G., {de Colle}, F., \&
  {V{\'a}zquez-Semadeni}, E. 2011, \mnras, 414, 1747,
  \dodoi{10.1111/j.1365-2966.2011.18507.x}

\bibitem[{{Ballesteros-Paredes}
  {et~al.}(1999{\natexlab{a}}){Ballesteros-Paredes}, {Hartmann}, \&
  {V{\'a}zquez-Semadeni}}]{Ballesteros-Paredesetal1999}
{Ballesteros-Paredes}, J., {Hartmann}, L., \& {V{\'a}zquez-Semadeni}, E.
  1999{\natexlab{a}}, \apj, 527, 285, \dodoi{10.1086/308076}

\bibitem[{{Ballesteros-Paredes}
  {et~al.}(1999{\natexlab{b}}){Ballesteros-Paredes}, {V{\'a}zquez-Semadeni}, \&
  {Scalo}}]{Ballesteros-Paredesetal1999b}
{Ballesteros-Paredes}, J., {V{\'a}zquez-Semadeni}, E., \& {Scalo}, J.
  1999{\natexlab{b}}, \apj, 515, 286, \dodoi{10.1086/307007}

\bibitem[{{Baug} {et~al.}(2020){Baug}, {Wang}, {Liu}, {Tang}, {Zhang}, {Li},
  {Eswaraiah}, {Liu}, {Tej}, {Goldsmith}, {Bronfman}, {Qin}, {T{\'o}th}, {Li},
  \& {Kim}}]{Baugetal2020}
{Baug}, T., {Wang}, K., {Liu}, T., {et~al.} 2020, \apj, 890, 44,
  \dodoi{10.3847/1538-4357/ab66b6}

\bibitem[{{Berry} {et~al.}(2005){Berry}, {Gledhill}, {Greaves}, \&
  {Jenness}}]{Berryetal2005}
{Berry}, D.~S., {Gledhill}, T.~M., {Greaves}, J.~S., \& {Jenness}, T. 2005, in
  Astronomical Society of the Pacific Conference Series, Vol. 343, Astronomical
  Polarimetry: Current Status and Future Directions, ed. A.~{Adamson},
  C.~{Aspin}, C.~{Davis}, \& T.~{Fujiyoshi}, 71

\bibitem[{{Bracco} {et~al.}(2017){Bracco}, {Palmeirim}, {Andr{\'e}}, {Adam},
  {Ade}, {Bacmann}, {Beelen}, {Beno{\^\i}t}, {Bideaud}, {Billot}, {Bourrion},
  {Calvo}, {Catalano}, {Coiffard}, {Comis}, {D'Addabbo}, {D{\'e}sert},
  {Didelon}, {Doyle}, {Goupy}, {K{\"o}nyves}, {Kramer}, {Lagache}, {Leclercq},
  {Mac{\'\i}as-P{\'e}rez}, {Maury}, {Mauskopf}, {Mayet}, {Monfardini}, {Motte},
  {Pajot}, {Pascale}, {Peretto}, {Perotto}, {Pisano}, {Ponthieu},
  {Rev{\'e}ret}, {Rigby}, {Ritacco}, {Rodriguez}, {Romero}, {Roy}, {Ruppin},
  {Schuster}, {Sievers}, {Triqueneaux}, {Tucker}, \& {Zylka}}]{Braccoetal2017}
{Bracco}, A., {Palmeirim}, P., {Andr{\'e}}, P., {et~al.} 2017, \aap, 604, A52,
  \dodoi{10.1051/0004-6361/201731117}

\bibitem[{{Chandrasekhar} \& {Fermi}(1953)}]{ChandrasekharFermi1953}
{Chandrasekhar}, S., \& {Fermi}, E. 1953, \apj, 118, 113,
  \dodoi{10.1086/145731}

\bibitem[{{Chapin} {et~al.}(2013){Chapin}, {Berry}, {Gibb}, {Jenness}, {Scott},
  {Tilanus}, {Economou}, \& {Holland}}]{Chapinetal2013}
{Chapin}, E.~L., {Berry}, D.~S., {Gibb}, A.~G., {et~al.} 2013, \mnras, 430,
  2545, \dodoi{10.1093/mnras/stt052}

\bibitem[{{Chapman} {et~al.}(2011){Chapman}, {Goldsmith}, {Pineda}, {Clemens},
  {Li}, \& {Kr{\v{c}}o}}]{Chapmanetal2011}
{Chapman}, N.~L., {Goldsmith}, P.~F., {Pineda}, J.~L., {et~al.} 2011, \apj,
  741, 21, \dodoi{10.1088/0004-637X/741/1/21}

\bibitem[{{Ching} {et~al.}(2017){Ching}, {Lai}, {Zhang}, {Girart}, {Qiu}, \&
  {Liu}}]{Chingetal2017}
{Ching}, T.-C., {Lai}, S.-P., {Zhang}, Q., {et~al.} 2017, \apj, 838, 121,
  \dodoi{10.3847/1538-4357/aa65cc}

\bibitem[{{Ching} {et~al.}(2018){Ching}, {Lai}, {Zhang}, {Girart}, {Qiu}, \&
  {Liu}}]{Chingetal2018}
---. 2018, \apj, 865, 110, \dodoi{10.3847/1538-4357/aad9fc}

\bibitem[{{Coud{\'e}} {et~al.}(2019){Coud{\'e}}, {Bastien}, {Houde}, {Sadavoy},
  {Friesen}, {Di Francesco}, {Johnstone}, {Mairs}, {Hasegawa}, {Kwon}, {Lai},
  {Qiu}, {Ward-Thompson}, {Berry}, {Chen}, {Fiege}, {Franzmann}, {Hatchell},
  {Lacaille}, {Matthews}, {Moriarty-Schieven}, {Pon}, {Andr{\'e}},
  {Arzoumanian}, {Aso}, {Byun}, {Eswaraiah}, {Chen}, {Chen}, {Ching}, {Cho},
  {Choi}, {Chrysostomou}, {Chung}, {Doi}, {Drabek-Maunder}, {Dowell}, {Eyres},
  {Falle}, {Friberg}, {Fuller}, {Furuya}, {Gledhill}, {Graves}, {Greaves},
  {Griffin}, {Gu}, {Hayashi}, {Hoang}, {Holland }, {Inoue}, {Inutsuka},
  {Iwasaki}, {Jeong}, {Kanamori}, {Kataoka}, {Kang}, {Kang}, {Kang},
  {Kawabata}, {Kemper}, {Kim}, {Kim}, {Kim}, {Kim}, {Kim}, {Kim}, {Kirk},
  {Kobayashi}, {Koch}, {Kwon}, {Lee}, {Lee}, {Lee}, {Li}, {Li}, {Li}, {Liu},
  {Liu}, {Liu}, {Liu}, {van Loo}, {Lyo}, {Matsumura}, {Nagata}, {Nakamura},
  {Nakanishi}, {Ohashi}, {Onaka}, {Parsons}, {Pattle}, {Peretto}, {Pyo},
  {Qian}, {Rao}, {Rawlings}, {Retter}, {Richer}, {Rigby}, {Robitaille},
  {Saito}, {Savini}, {Scaife}, {Seta}, {Shinnaga}, {Soam}, {Tamura}, {Tang},
  {Tomisaka}, {Tsukamoto}, {Wang}, {Wang}, {Whitworth}, {Yen}, {Yoo}, {Yuan},
  {Zenko}, {Zhang}, {Zhang}, {Zhou}, {Zhu}, \& fields In STar-forming Regions
  Observations (BISTRO~Collaboration)}]{Coudeetal2019}
{Coud{\'e}}, S., {Bastien}, P., {Houde}, M., {et~al.} 2019, \apj, 877, 88,
  \dodoi{10.3847/1538-4357/ab1b23}

\bibitem[{{Crutcher}(2012)}]{Crutcher2012}
{Crutcher}, R.~M. 2012, \araa, 50, 29,
  \dodoi{10.1146/annurev-astro-081811-125514}

\bibitem[{{Crutcher} {et~al.}(2004){Crutcher}, {Nutter}, {Ward-Thompson}, \&
  {Kirk}}]{Crutcheretal2004}
{Crutcher}, R.~M., {Nutter}, D.~J., {Ward-Thompson}, D., \& {Kirk}, J.~M. 2004,
  \apj, 600, 279, \dodoi{10.1086/379705}

\bibitem[{{Curran} \& {Chrysostomou}(2007)}]{CurranChrysostomou2007}
{Curran}, R.~L., \& {Chrysostomou}, A. 2007, \mnras, 382, 699,
  \dodoi{10.1111/j.1365-2966.2007.12399.x}

\bibitem[{{Davidson} {et~al.}(2011){Davidson}, {Novak}, {Matthews}, {Matthews},
  {Goldsmith}, {Chapman}, {Volgenau}, {Vaillancourt}, \&
  {Attard}}]{Davidsonetal2011}
{Davidson}, J.~A., {Novak}, G., {Matthews}, T.~G., {et~al.} 2011, \apj, 732,
  97, \dodoi{10.1088/0004-637X/732/2/97}

\bibitem[{{Davis} {et~al.}(2010){Davis}, {Chrysostomou}, {Hatchell},
  {Wouterloot}, {Buckle}, {Nutter}, {Fich}, {Brunt}, {Butner}, {Cavanagh},
  {Curtis}, {Duarte-Cabral}, {di Francesco}, {Etxaluze}, {Friberg}, {Friesen},
  {Fuller}, {Graves}, {Greaves}, {Hogerheijde}, {Johnstone}, {Matthews},
  {Matthews}, {Rawlings}, {Richer}, {Roberts}, {Sadavoy}, {Simpson}, {Tothill},
  {Tsamis}, {Viti}, {Ward-Thompson}, {White}, \& {Yates}}]{Davisetal2010}
{Davis}, C.~J., {Chrysostomou}, A., {Hatchell}, J., {et~al.} 2010, \mnras, 405,
  759, \dodoi{10.1111/j.1365-2966.2010.16499.x}

\bibitem[{{Davis}(1951)}]{Davis1951}
{Davis}, L. 1951, Physical Review, 81, 890, \dodoi{10.1103/PhysRev.81.890.2}

\bibitem[{{Dempsey} {et~al.}(2013){Dempsey}, {Friberg}, {Jenness}, {Tilanus},
  {Thomas}, {Holland}, {Bintley}, {Berry}, {Chapin}, {Chrysostomou}, {Davis},
  {Gibb}, {Parsons}, \& {Robson}}]{Dempseyetal2013}
{Dempsey}, J.~T., {Friberg}, P., {Jenness}, T., {et~al.} 2013, \mnras, 430,
  2534, \dodoi{10.1093/mnras/stt090}

\bibitem[{{Doi} {et~al.}(2020){Doi}, {Hasegawa}, {Furuya}, {Coud{\'e}}, {Hull},
  {Arzoumanian}, {Bastien}, {Chen}, {Di Francesco}, {Friesen}, {Houde},
  {Inutsuka}, {Mairs}, {Matsumura}, {Onaka}, {Sadavoy}, {Shimajiri}, {Tahani},
  {Tomisaka}, {Eswaraiah}, {Koch}, {Pattle}, {Won Lee}, {Tamura}, {Berry},
  {Ching}, {Hwang}, {Kwon}, {Soam}, {Wang}, {Lai}, {Qiu}, {Ward-Thompson},
  {Byun}, {Chen}, {Chen}, {Chen}, {Cho}, {Choi}, {Choi}, {Chrysostomou},
  {Chung}, {Diep}, {Duan}, {Fanciullo}, {Fiege}, {Franzmann}, {Friberg},
  {Fuller}, {Gledhill}, {Graves}, {Greaves}, {Griffin}, {Gu}, {Han},
  {Hatchell}, {Hayashi}, {Hoang}, {Inoue}, {Iwasaki}, {Jeong}, {Johnstone},
  {Kanamori}, {Kang}, {Kang}, {Kang}, {Kataoka}, {Kawabata}, {Kemper}, {Kim},
  {Kim}, {Kim}, {Kim}, {Kim}, {Kim}, {Kirk}, {Kobayashi}, {Konyves}, {Kusune},
  {Kwon}, {Lacaille}, {Law}, {Lee}, {Lee}, {Lee}, {Lee}, {Lee}, {Li}, {Li},
  {Li}, {Liu}, {Liu}, {Liu}, {Liu}, {de Looze}, {Lyo}, {Matthews},
  {Moriarty-Schieven}, {Nagata}, {Nakamura}, {Nakanishi}, {Ohashi}, {Park},
  {Parsons}, {Peretto}, {Pyo}, {Qian}, {Rao}, {Rawlings}, {Retter}, {Richer},
  {Rigby}, {Saito}, {Savini}, {Scaife}, {Seta}, {Shinnaga}, {Tang},
  {Tsukamoto}, {Viti}, {Wang}, {Whitworth}, {Yen}, {Yoo}, {Yuan}, {Yun},
  {Zenko}, {Zhang}, {Zhang}, {Zhang}, {Zhou}, {Zhu}, {Andr{\'e}}, {Dowell},
  {Eyres}, {Falle}, {van Loo}, \& {Robitaille}}]{Doietal2020}
{Doi}, Y., {Hasegawa}, T., {Furuya}, R.~S., {et~al.} 2020, \apj, 899, 28,
  \dodoi{10.3847/1538-4357/aba1e2}

\bibitem[{{Elias}(1978)}]{Elias1978}
{Elias}, J.~H. 1978, \apj, 224, 857, \dodoi{10.1086/156436}

\bibitem[{{Eswaraiah} {et~al.}(2020){Eswaraiah}, {Li}, {Samal}, {Wang}, {Ma},
  {Lai}, {Zavagno}, {Ching}, {Liu}, {Pattle}, {Ward-Thompson}, {Pandey}, \&
  {Ojha}}]{Eswaraiahetal2020}
{Eswaraiah}, C., {Li}, D., {Samal}, M.~R., {et~al.} 2020, \apj, 897, 90,
  \dodoi{10.3847/1538-4357/ab83f2}

\bibitem[{{Fiedler} \& {Mouschovias}(1992)}]{FiedlerMouschovias1992}
{Fiedler}, R.~A., \& {Mouschovias}, T.~C. 1992, \apj, 391, 199,
  \dodoi{10.1086/171336}

\bibitem[{{Fiedler} \& {Mouschovias}(1993)}]{FiedlerMouschovias1993}
---. 1993, \apj, 415, 680, \dodoi{10.1086/173193}

\bibitem[{{Friberg} {et~al.}(2016){Friberg}, {Bastien}, {Berry}, {Savini},
  {Graves}, \& {Pattle}}]{Fribergetal2016}
{Friberg}, P., {Bastien}, P., {Berry}, D., {et~al.} 2016, in Society of
  Photo-Optical Instrumentation Engineers (SPIE) Conference Series, Vol. 9914,
  \procspie, 991403, \dodoi{10.1117/12.2231943}

\bibitem[{{Friberg} {et~al.}(2018){Friberg}, {Berry}, {Savini}, {Bintley},
  {Dempsey}, {Graves}, \& {Parsons}}]{Fribergetal2018}
{Friberg}, P., {Berry}, D., {Savini}, G., {et~al.} 2018, in Society of
  Photo-Optical Instrumentation Engineers (SPIE) Conference Series, Vol. 10708,
  \procspie, 107083M, \dodoi{10.1117/12.2314345}

\bibitem[{{Galli} \& {Shu}(1993)}]{GalliShu1993}
{Galli}, D., \& {Shu}, F.~H. 1993, \apj, 417, 220, \dodoi{10.1086/173305}

\bibitem[{{Girart} {et~al.}(2006){Girart}, {Rao}, \&
  {Marrone}}]{Girartetal2006}
{Girart}, J.~M., {Rao}, R., \& {Marrone}, D.~P. 2006, Science, 313, 812,
  \dodoi{10.1126/science.1129093}

\bibitem[{{Goodman} {et~al.}(1990){Goodman}, {Bastien}, {Myers}, \&
  {Menard}}]{Goodmanetal1990}
{Goodman}, A.~A., {Bastien}, P., {Myers}, P.~C., \& {Menard}, F. 1990, \apj,
  359, 363, \dodoi{10.1086/169070}

\bibitem[{{Goodman} {et~al.}(1993){Goodman}, {Benson}, {Fuller}, \&
  {Myers}}]{Goodmanetal1993}
{Goodman}, A.~A., {Benson}, P.~J., {Fuller}, G.~A., \& {Myers}, P.~C. 1993,
  \apj, 406, 528, \dodoi{10.1086/172465}

\bibitem[{{Goodman} {et~al.}(1992){Goodman}, {Jones}, {Lada}, \&
  {Myers}}]{Goodmanetal1992}
{Goodman}, A.~A., {Jones}, T.~J., {Lada}, E.~A., \& {Myers}, P.~C. 1992, \apj,
  399, 108, \dodoi{10.1086/171907}

\bibitem[{{Hartmann}(2002)}]{Hartmann2002}
{Hartmann}, L. 2002, \apj, 578, 914, \dodoi{10.1086/342657}

\bibitem[{{Hartmann} {et~al.}(2001){Hartmann}, {Ballesteros-Paredes}, \&
  {Bergin}}]{Hartmannetal2001}
{Hartmann}, L., {Ballesteros-Paredes}, J., \& {Bergin}, E.~A. 2001, \apj, 562,
  852, \dodoi{10.1086/323863}

\bibitem[{{Heiles}(2000)}]{Heiles2000}
{Heiles}, C. 2000, \aj, 119, 923, \dodoi{10.1086/301236}

\bibitem[{{Henshaw} {et~al.}(2016){Henshaw}, {Longmore}, {Kruijssen}, {Davies},
  {Bally}, {Barnes}, {Battersby}, {Burton}, {Cunningham}, {Dale}, {Ginsburg},
  {Immer}, {Jones}, {Kendrew}, {Mills}, {Molinari}, {Moore}, {Ott}, {Pillai},
  {Rathborne}, {Schilke}, {Schmiedeke}, {Testi}, {Walker}, {Walsh}, \&
  {Zhang}}]{Henshawetal2016}
{Henshaw}, J.~D., {Longmore}, S.~N., {Kruijssen}, J.~M.~D., {et~al.} 2016,
  \mnras, 457, 2675, \dodoi{10.1093/mnras/stw121}

\bibitem[{{Heyer} {et~al.}(1987){Heyer}, {Vrba}, {Snell}, {Schloerb}, {Strom},
  {Goldsmith}, \& {Strom}}]{Heyeretal1987}
{Heyer}, M.~H., {Vrba}, F.~J., {Snell}, R.~L., {et~al.} 1987, \apj, 321, 855,
  \dodoi{10.1086/165678}

\bibitem[{{Hildebrand}(1983)}]{Hildebrand1983}
{Hildebrand}, R.~H. 1983, \qjras, 24, 267

\bibitem[{{Hildebrand} {et~al.}(2009){Hildebrand}, {Kirby}, {Dotson}, {Houde},
  \& {Vaillancourt}}]{Hildebrandetal2009}
{Hildebrand}, R.~H., {Kirby}, L., {Dotson}, J.~L., {Houde}, M., \&
  {Vaillancourt}, J.~E. 2009, \apj, 696, 567,
  \dodoi{10.1088/0004-637X/696/1/567}

\bibitem[{{Holland} {et~al.}(2013){Holland}, {Bintley}, {Chapin},
  {Chrysostomou}, {Davis}, {Dempsey}, {Duncan}, {Fich}, {Friberg}, {Halpern},
  {Irwin}, {Jenness}, {Kelly}, {MacIntosh}, {Robson}, {Scott}, {Ade},
  {Atad-Ettedgui}, {Berry}, {Craig}, {Gao}, {Gibb}, {Hilton}, {Hollister},
  {Kycia}, {Lunney}, {McGregor}, {Montgomery}, {Parkes}, {Tilanus}, {Ullom},
  {Walther}, {Walton}, {Woodcraft}, {Amiri}, {Atkinson}, {Burger}, {Chuter},
  {Coulson}, {Doriese}, {Dunare}, {Economou}, {Niemack}, {Parsons},
  {Reintsema}, {Sibthorpe}, {Smail}, {Sudiwala}, \& {Thomas}}]{Hollandetal2013}
{Holland}, W.~S., {Bintley}, D., {Chapin}, E.~L., {et~al.} 2013, \mnras, 430,
  2513, \dodoi{10.1093/mnras/sts612}

\bibitem[{{Hull} \& {Zhang}(2019)}]{HullZhang2019}
{Hull}, C. L.~H., \& {Zhang}, Q. 2019, Frontiers in Astronomy and Space
  Sciences, 6, 3, \dodoi{10.3389/fspas.2019.00003}

\bibitem[{{Hull} {et~al.}(2013){Hull}, {Plambeck}, {Bolatto}, {Bower},
  {Carpenter}, {Crutcher}, {Fiege}, {Franzmann}, {Hakobian}, {Heiles}, {Houde},
  {Hughes}, {Jameson}, {Kwon}, {Lamb}, {Looney}, {Matthews}, {Mundy}, {Pillai},
  {Pound}, {Stephens}, {Tobin}, {Vaillancourt}, {Volgenau}, \&
  {Wright}}]{Hulletal2013}
{Hull}, C. L.~H., {Plambeck}, R.~L., {Bolatto}, A.~D., {et~al.} 2013, \apj,
  768, 159, \dodoi{10.1088/0004-637X/768/2/159}

\bibitem[{{Hull} {et~al.}(2014){Hull}, {Plambeck}, {Kwon}, {Bower},
  {Carpenter}, {Crutcher}, {Fiege}, {Franzmann}, {Hakobian}, {Heiles}, {Houde},
  {Hughes}, {Lamb}, {Looney}, {Marrone}, {Matthews}, {Pillai}, {Pound},
  {Rahman}, {Sandell}, {Stephens}, {Tobin}, {Vaillancourt}, {Volgenau}, \&
  {Wright}}]{Hulletal2014}
{Hull}, C. L.~H., {Plambeck}, R.~L., {Kwon}, W., {et~al.} 2014, \apjs, 213, 13,
  \dodoi{10.1088/0067-0049/213/1/13}

\bibitem[{{Hull} {et~al.}(2017{\natexlab{a}}){Hull}, {Mocz}, {Burkhart},
  {Goodman}, {Girart}, {Cort{\'e}s}, {Hernquist}, {Springel}, {Li}, \&
  {Lai}}]{Hulletal2017a}
{Hull}, C. L.~H., {Mocz}, P., {Burkhart}, B., {et~al.} 2017{\natexlab{a}},
  \apjl, 842, L9, \dodoi{10.3847/2041-8213/aa71b7}

\bibitem[{{Hull} {et~al.}(2017{\natexlab{b}}){Hull}, {Girart}, {Tychoniec},
  {Rao}, {Cort{\'e}s}, {Pokhrel}, {Zhang}, {Houde}, {Dunham}, {Kristensen},
  {Lai}, {Li}, \& {Plambeck}}]{Hulletal2017b}
{Hull}, C. L.~H., {Girart}, J.~M., {Tychoniec}, {\L}., {et~al.}
  2017{\natexlab{b}}, \apj, 847, 92, \dodoi{10.3847/1538-4357/aa7fe9}

\bibitem[{{Johnstone} {et~al.}(2017){Johnstone}, {Ciccone}, {Kirk}, {Mairs},
  {Buckle}, {Berry}, {Broekhoven-Fiene}, {Currie}, {Hatchell}, {Jenness},
  {Mottram}, {Pattle}, {Tisi}, {Di Francesco}, {Hogerheijde}, {Ward-Thompson},
  {Bastien}, {Bresnahan}, {Butner}, {Chen}, {Chrysostomou}, {Coud{\'e}},
  {Davis}, {Drabek-Maunder}, {Duarte-Cabral}, {Fich}, {Fiege}, {Friberg},
  {Friesen}, {Fuller}, {Graves}, {Greaves}, {Gregson}, {Holland}, {Joncas},
  {Kirk}, {Knee}, {Marsh}, {Matthews}, {Moriarty-Schieven}, {Mowat}, {Nutter},
  {Pineda}, {Salji}, {Rawlings}, {Richer}, {Robertson}, {Rosolowsky}, {Rumble},
  {Sadavoy}, {Thomas}, {Tothill}, {Viti}, {White}, {Wouterloot}, {Yates}, \&
  {Zhu}}]{Johnstoneetal2017}
{Johnstone}, D., {Ciccone}, S., {Kirk}, H., {et~al.} 2017, \apj, 836, 132,
  \dodoi{10.3847/1538-4357/aa5b95}

\bibitem[{{Kenyon} {et~al.}(1993){Kenyon}, {Calvet}, \&
  {Hartmann}}]{Kenyonetal1993}
{Kenyon}, S.~J., {Calvet}, N., \& {Hartmann}, L. 1993, \apj, 414, 676,
  \dodoi{10.1086/173114}

\bibitem[{{Kenyon} {et~al.}(1990){Kenyon}, {Hartmann}, {Strom}, \&
  {Strom}}]{Kenyonetal1990}
{Kenyon}, S.~J., {Hartmann}, L.~W., {Strom}, K.~M., \& {Strom}, S.~E. 1990,
  \aj, 99, 869, \dodoi{10.1086/115380}

\bibitem[{{Kwon} {et~al.}(2018){Kwon}, {Doi}, {Tamura}, {Matsumura}, {Pattle},
  {Berry}, {Sadavoy}, {Matthews}, {Ward-Thompson}, {Hasegawa}, {Furuya}, {Pon},
  {Di Francesco}, {Arzoumanian}, {Hayashi}, {Kawabata}, {Onaka}, {Choi},
  {Kang}, {Hoang}, {Lee}, {Lee}, {Liu}, {Liu}, {Inutsuka}, {Eswaraiah},
  {Bastien}, {Kwon}, {Lai}, {Qiu}, {Coud{\'e}}, {Franzmann}, {Friberg},
  {Graves}, {Greaves}, {Houde}, {Johnstone}, {Kirk}, {Koch}, {Li}, {Parsons},
  {Rao}, {Rawlings}, {Shinnaga}, {van Loo}, {Aso}, {Byun}, {Chen}, {Chen},
  {Chen}, {Ching}, {Cho}, {Chrysostomou}, {Chung}, {Drabek-Maunder}, {Eyres},
  {Fiege}, {Friesen}, {Fuller}, {Gledhill}, {Griffin}, {Gu}, {Hatchell},
  {Holland }, {Inoue}, {Iwasaki}, {Jeong}, {Kang}, {Kang}, {Kemper}, {Kim},
  {Kim}, {Kim}, {Kim}, {Kim}, {Kim}, {Lacaille}, {Lee}, {Li}, {Li}, {Liu},
  {Liu}, {Lyo}, {Mairs}, {Moriarty-Schieven}, {Nakamura}, {Nakanishi},
  {Ohashi}, {Peretto}, {Pyo}, {Qian}, {Retter}, {Richer}, {Rigby},
  {Robitaille}, {Savini}, {Scaife}, {Soam}, {Tang}, {Tomisaka}, {Wang}, {Wang},
  {Whitworth}, {Yen}, {Yoo}, {Yuan}, {Zhang}, {Zhang}, {Zhou}, {Zhu},
  {Andr{\'e}}, {Dowell}, {Falle}, {Tsukamoto}, {Nakagawa}, {Kanamori},
  {Kataoka}, {Kobayashi}, {Nagata}, {Saito}, {Seta}, \& {Zenko}}]{Kwonetal2018}
{Kwon}, J., {Doi}, Y., {Tamura}, M., {et~al.} 2018, \apj, 859, 4,
  \dodoi{10.3847/1538-4357/aabd82}

\bibitem[{{Li} {et~al.}(2006){Li}, {Griffin}, {Krejny}, {Novak}, {Loewenstein},
  {Newcomb}, {Calisse}, \& {Chuss}}]{LiHBetal2006}
{Li}, H., {Griffin}, G.~S., {Krejny}, M., {et~al.} 2006, \apj, 648, 340,
  \dodoi{10.1086/505858}

\bibitem[{{Li} {et~al.}(2009){Li}, {Dowell}, {Goodman}, {Hildebrand }, \&
  {Novak}}]{LiHBetal2009}
{Li}, H.-b., {Dowell}, C.~D., {Goodman}, A., {Hildebrand }, R., \& {Novak}, G.
  2009, \apj, 704, 891, \dodoi{10.1088/0004-637X/704/2/891}

\bibitem[{{Li} {et~al.}(2014){Li}, {Krasnopolsky}, {Shang}, \&
  {Zhao}}]{LiZYetal2014}
{Li}, Z.-Y., {Krasnopolsky}, R., {Shang}, H., \& {Zhao}, B. 2014, \apj, 793,
  130, \dodoi{10.1088/0004-637X/793/2/130}

\bibitem[{{Liu} {et~al.}(2020){Liu}, {Zhang}, {Qiu}, {Liu}, {Pillai}, {Girart},
  {Li}, \& {Wang}}]{LiuJunhaoetal2020}
{Liu}, J., {Zhang}, Q., {Qiu}, K., {et~al.} 2020, \apj, 895, 142,
  \dodoi{10.3847/1538-4357/ab9087}

\bibitem[{{Liu} {et~al.}(2019){Liu}, {Qiu}, {Berry}, {Di Francesco}, {Bastien},
  {Koch}, {Furuya}, {Kim}, {Coud{\'e}}, {Lee}, {Soam}, {Eswaraiah}, {Li},
  {Hwang}, {Lyo}, {Pattle}, {Hasegawa}, {Kwon}, {Lai}, {Ward-Thompson},
  {Ching}, {Chen}, {Gu}, {Li}, {Li}, {Liu}, {Qian}, {Wang}, {Yuan}, {Zhang},
  {Zhang}, {Zhang}, {Zhou}, {Zhu}, {Andr{\'e}}, {Arzoumanian}, {Aso}, {Byun},
  {Chen}, {Chen}, {Chen}, {Cho}, {Choi}, {Chrysostomou}, {Chung}, {Doi},
  {Drabek-Maunder}, {Dowell}, {Eyres}, {Falle}, {Fanciullo}, {Fiege},
  {Franzmann}, {Friberg}, {Friesen}, {Fuller}, {Gledhill}, {Graves}, {Greaves},
  {Griffin}, {Han}, {Hatchell}, {Hayashi}, {Hoang}, {Holland}, {Houde},
  {Inoue}, {Inutsuka}, {Iwasaki}, {Jeong}, {Johnstone}, {Kanamori}, {Kang},
  {Kang}, {Kang}, {Kataoka}, {Kawabata}, {Kemper}, {Kim}, {Kim}, {Kim}, {Kim},
  {Kim}, {Kirk}, {Kobayashi}, {Kusune}, {Kwon}, {Lacaille}, {Lee}, {Lee},
  {Lee}, {Lee}, {Liu}, {Liu}, {van Loo}, {Mairs}, {Matsumura}, {Matthews},
  {Moriarty-Schieven}, {Nagata}, {Nakamura}, {Nakanishi}, {Ohashi}, {Onaka},
  {Parker}, {Parsons}, {Pascale}, {Peretto}, {Pon}, {Pyo}, {Rao}, {Rawlings},
  {Retter}, {Richer}, {Rigby}, {Robitaille}, {Sadavoy}, {Saito}, {Savini},
  {Scaife}, {Seta}, {Shinnaga}, {Tamura}, {Tang}, {Tomisaka}, {Tsukamoto},
  {Wang}, {Whitworth}, {Yen}, {Yoo}, \& {Zenko}}]{LiuJunhaoetal2019}
{Liu}, J., {Qiu}, K., {Berry}, D., {et~al.} 2019, \apj, 877, 43,
  \dodoi{10.3847/1538-4357/ab0958}

\bibitem[{{Mac Low} \& {Klessen}(2004)}]{MacLowKlessen2004}
{Mac Low}, M.-M., \& {Klessen}, R.~S. 2004, Reviews of Modern Physics, 76, 125,
  \dodoi{10.1103/RevModPhys.76.125}

\bibitem[{{Marsh} {et~al.}(2016){Marsh}, {Kirk}, {Andr{\'e}}, {Griffin},
  {K{\"o}nyves}, {Palmeirim}, {Men'shchikov}, {Ward-Thompson}, {Benedettini},
  {Bresnahan}, {di Francesco}, {Elia}, {Motte}, {Peretto}, {Pezzuto}, {Roy},
  {Sadavoy}, {Schneider}, {Spinoglio}, \& {White}}]{Marshetal2016}
{Marsh}, K.~A., {Kirk}, J.~M., {Andr{\'e}}, P., {et~al.} 2016, \mnras, 459,
  342, \dodoi{10.1093/mnras/stw301}

\bibitem[{{McKee} \& {Ostriker}(2007)}]{McKeeOstriker2007}
{McKee}, C.~F., \& {Ostriker}, E.~C. 2007, \araa, 45, 565,
  \dodoi{10.1146/annurev.astro.45.051806.110602}

\bibitem[{{Mestel} \& {Spitzer}(1956)}]{MestelSpitzer1956}
{Mestel}, L., \& {Spitzer}, L., J. 1956, \mnras, 116, 503,
  \dodoi{10.1093/mnras/116.5.503}

\bibitem[{{Mizuno} {et~al.}(1994){Mizuno}, {Onishi}, {Hayashi}, {Ohashi},
  {Sunada}, {Hasegawa}, \& {Fukui}}]{Mizunoetal1994}
{Mizuno}, A., {Onishi}, T., {Hayashi}, M., {et~al.} 1994, \nat, 368, 719,
  \dodoi{10.1038/368719a0}

\bibitem[{{Mocz} {et~al.}(2017){Mocz}, {Burkhart}, {Hernquist}, {McKee}, \&
  {Springel}}]{Moczetal2017}
{Mocz}, P., {Burkhart}, B., {Hernquist}, L., {McKee}, C.~F., \& {Springel}, V.
  2017, \apj, 838, 40, \dodoi{10.3847/1538-4357/aa6475}

\bibitem[{{Mouschovias}(1991)}]{Mouschovias1991}
{Mouschovias}, T.~C. 1991, \apj, 373, 169, \dodoi{10.1086/170035}

\bibitem[{{Mouschovias} \& {Spitzer}(1976)}]{MouschoviasSpitzer1976}
{Mouschovias}, T.~C., \& {Spitzer}, L., J. 1976, \apj, 210, 326,
  \dodoi{10.1086/154835}

\bibitem[{{Mouschovias} {et~al.}(2006){Mouschovias}, {Tassis}, \&
  {Kunz}}]{Mouschoviasetal2006}
{Mouschovias}, T.~C., {Tassis}, K., \& {Kunz}, M.~W. 2006, \apj, 646, 1043,
  \dodoi{10.1086/500125}

\bibitem[{{Nakano} \& {Nakamura}(1978)}]{NakanoNakamura1978}
{Nakano}, T., \& {Nakamura}, T. 1978, \pasj, 30, 671

\bibitem[{{Ohashi} {et~al.}(1997){Ohashi}, {Hayashi}, {Ho}, {Momose}, {Tamura},
  {Hirano}, \& {Sargent}}]{Ohashietal1997}
{Ohashi}, N., {Hayashi}, M., {Ho}, P. T.~P., {et~al.} 1997, \apj, 488, 317,
  \dodoi{10.1086/304685}

\bibitem[{{Ostriker} {et~al.}(2001){Ostriker}, {Stone}, \&
  {Gammie}}]{Ostrikeretal2001}
{Ostriker}, E.~C., {Stone}, J.~M., \& {Gammie}, C.~F. 2001, \apj, 546, 980,
  \dodoi{10.1086/318290}

\bibitem[{{Palmeirim} {et~al.}(2013){Palmeirim}, {Andr{\'e}}, {Kirk},
  {Ward-Thompson}, {Arzoumanian}, {K{\"o}nyves}, {Didelon}, {Schneider},
  {Benedettini}, {Bontemps}, {Di Francesco}, {Elia}, {Griffin}, {Hennemann},
  {Hill}, {Martin}, {Men'shchikov}, {Molinari}, {Motte}, {Nguyen Luong},
  {Nutter}, {Peretto}, {Pezzuto}, {Roy}, {Rygl}, {Spinoglio}, \&
  {White}}]{Palmeirimetal2013}
{Palmeirim}, P., {Andr{\'e}}, P., {Kirk}, J., {et~al.} 2013, \aap, 550, A38,
  \dodoi{10.1051/0004-6361/201220500}

\bibitem[{{Pattle} {et~al.}(2017){Pattle}, {Ward-Thompson}, {Berry},
  {Hatchell}, {Chen}, {Pon}, {Koch}, {Kwon}, {Kim}, {Bastien}, {Cho},
  {Coud{\'e}}, {Di Francesco}, {Fuller}, {Furuya}, {Graves}, {Johnstone},
  {Kirk}, {Kwon}, {Lee}, {Matthews}, {Mottram}, {Parsons}, {Sadavoy},
  {Shinnaga}, {Soam}, {Hasegawa}, {Lai}, {Qiu}, \& {Friberg}}]{Pattleetal2017}
{Pattle}, K., {Ward-Thompson}, D., {Berry}, D., {et~al.} 2017, \apj, 846, 122,
  \dodoi{10.3847/1538-4357/aa80e5}

\bibitem[{{Pattle} {et~al.}(2018){Pattle}, {Ward-Thompson}, {Hasegawa},
  {Bastien}, {Kwon}, {Lai}, {Qiu}, {Furuya}, {Berry}, \& {The JCMT BISTRO
  Survey Team}}]{Pattleetal2018}
{Pattle}, K., {Ward-Thompson}, D., {Hasegawa}, T., {et~al.} 2018, \apjl, 860,
  L6, \dodoi{10.3847/2041-8213/aac771}

\bibitem[{{Pattle} {et~al.}(2019){Pattle}, {Lai}, {Hasegawa}, {Wang}, {Furuya},
  {Ward-Thompson}, {Bastien}, {Coud{\'e}}, {Eswaraiah}, {Fanciullo}, {di
  Francesco}, {Hoang}, {Kim}, {Kwon}, {Lee}, {Liu}, {Liu}, {Matsumura},
  {Onaka}, {Sadavoy}, \& {Soam}}]{Pattleetal2019}
{Pattle}, K., {Lai}, S.-P., {Hasegawa}, T., {et~al.} 2019, \apj, 880, 27,
  \dodoi{10.3847/1538-4357/ab286f}

\bibitem[{{Pattle} {et~al.}(2020){Pattle}, {Lai}, {Di Francesco}, {Sadavoy},
  {Ward-Thompson}, {Johnstone}, {Hoang}, {Arzoumanian}, {Bastien}, {Bourke},
  {Coud{\'e}}, {Doi}, {Eswaraiah}, {Fanciullo}, {Furuya}, {Hwang}, {Hull},
  {Kang}, {Kim}, {Kirchschlager}, {Kwon}, {Kwon}, {Lee}, {Liu}, {Redman},
  {Soam}, {Tahani}, {Tamura}, \& {Tang}}]{Pattleetal2020}
{Pattle}, K., {Lai}, S.-P., {Di Francesco}, J., {et~al.} 2020, arXiv e-prints,
  arXiv:2011.09765.
\newblock \doarXiv{2011.09765}

\bibitem[{{Pillai} {et~al.}(2020){Pillai}, {Clemens}, {Reissl}, {Myers},
  {Kauffmann}, {Lopez-Rodriguez}, {Alves}, {Franco}, {Henshaw}, {Menten},
  {Nakamura}, {Seifried}, {Sugitani}, \& {Wiesemeyer}}]{Pillaietal2020}
{Pillai}, T. G.~S., {Clemens}, D.~P., {Reissl}, S., {et~al.} 2020, Nature
  Astronomy, 4, 1195, \dodoi{10.1038/s41550-020-1172-6}

\bibitem[{{Planck Collaboration} {et~al.}(2015){Planck Collaboration}, {Ade},
  {Aghanim}, {Alina}, {Alves}, {Armitage-Caplan}, {Arnaud}, {Arzoumanian},
  {Ashdown}, {Atrio-Barand ela}, {Aumont}, {Baccigalupi}, {Banday}, {Barreiro},
  {Battaner}, {Benabed}, {Benoit-L{\'e}vy}, {Bernard}, {Bersanelli},
  {Bielewicz}, {Bock}, {Bond}, {Borrill}, {Bouchet}, {Boulanger}, {Bracco},
  {Burigana}, {Butler}, {Cardoso}, {Catalano}, {Chamballu}, {Chary}, {Chiang},
  {Christensen}, {Colombi}, {Colombo}, {Combet}, {Couchot}, {Coulais}, {Crill},
  {Curto}, {Cuttaia}, {Danese}, {Davies}, {Davis}, {de Bernardis}, {de Gouveia
  Dal Pino}, {de Rosa}, {de Zotti}, {Delabrouille}, {D{\'e}sert}, {Dickinson},
  {Diego}, {Donzelli}, {Dor{\'e}}, {Douspis}, {Dunkley}, {Dupac}, {Efstathiou},
  {En{\ss}lin}, {Eriksen}, {Falgarone}, {Ferri{\`e}re}, {Finelli}, {Forni},
  {Frailis}, {Fraisse}, {Franceschi}, {Galeotta}, {Ganga}, {Ghosh}, {Giard},
  {Giraud-H{\'e}raud}, {Gonz{\'a}lez-Nuevo}, {G{\'o}rski}, {Gregorio},
  {Gruppuso}, {Guillet}, {Hansen}, {Harrison}, {Helou},
  {Hern{\'a}ndez-Monteagudo}, {Hildebrand t}, {Hivon}, {Hobson}, {Holmes},
  {Hornstrup}, {Huffenberger}, {Jaffe}, {Jaffe}, {Jones}, {Juvela},
  {Keih{\"a}nen}, {Keskitalo}, {Kisner}, {Kneissl}, {Knoche}, {Kunz},
  {Kurki-Suonio}, {Lagache}, {L{\"a}hteenm{\"a}ki}, {Lamarre}, {Lasenby},
  {Lawrence}, {Leahy}, {Leonardi}, {Levrier}, {Liguori}, {Lilje},
  {Linden-V{\o}rnle}, {L{\'o}pez-Caniego}, {Lubin}, {Mac{\'\i}as-P{\'e}rez},
  {Maffei}, {Magalh{\~a}es}, {Maino}, {Mandolesi}, {Maris}, {Marshall},
  {Martin}, {Mart{\'\i}nez-Gonz{\'a}lez}, {Masi}, {Matarrese}, {Mazzotta},
  {Melchiorri}, {Mendes}, {Mennella}, {Migliaccio}, {Miville-Desch{\^e}nes},
  {Moneti}, {Montier}, {Morgante}, {Mortlock}, {Munshi}, {Murphy}, {Naselsky},
  {Nati}, {Natoli}, {Netterfield}, {Noviello}, {Novikov}, {Novikov},
  {Oxborrow}, {Pagano}, {Pajot}, {Paladini}, {Paoletti}, {Pasian}, {Pearson},
  {Perdereau}, {Perotto}, {Perrotta}, {Piacentini}, {Piat}, {Pietrobon},
  {Plaszczynski}, {Poidevin}, {Pointecouteau}, {Polenta}, {Popa}, {Pratt},
  {Prunet}, {Puget}, {Rachen}, {Reach}, {Rebolo}, {Reinecke}, {Remazeilles},
  {Renault}, {Ricciardi}, {Riller}, {Ristorcelli}, {Rocha}, {Rosset},
  {Roudier}, {Rubi{\~n}o-Mart{\'\i}n}, {Rusholme}, {Sandri}, {Savini}, {Scott},
  {Spencer}, {Stolyarov}, {Stompor}, {Sudiwala}, {Sutton}, {Suur-Uski},
  {Sygnet}, {Tauber}, {Terenzi}, {Toffolatti}, {Tomasi}, {Tristram}, {Tucci},
  {Umana}, {Valenziano}, {Valiviita}, {Van Tent}, {Vielva}, {Villa}, {Wade},
  {Wandelt}, {Zacchei}, \& {Zonca}}]{PlanckCollaborationXIX2015}
{Planck Collaboration}, {Ade}, P.~A.~R., {Aghanim}, N., {et~al.} 2015, \aap,
  576, A104, \dodoi{10.1051/0004-6361/201424082}

\bibitem[{{Planck Collaboration} {et~al.}(2016{\natexlab{a}}){Planck
  Collaboration}, {Ade}, {Aghanim}, {Alves}, {Arnaud}, {Arzoumanian},
  {Ashdown}, {Aumont}, {Baccigalupi}, {Band ay}, {Barreiro}, {Bartolo},
  {Battaner}, {Benabed}, {Beno{\^\i}t}, {Benoit-L{\'e}vy}, {Bernard},
  {Bersanelli}, {Bielewicz}, {Bock}, {Bonavera}, {Bond}, {Borrill}, {Bouchet},
  {Boulanger}, {Bracco}, {Burigana}, {Calabrese}, {Cardoso}, {Catalano},
  {Chiang}, {Christensen}, {Colombo}, {Combet}, {Couchot}, {Crill}, {Curto},
  {Cuttaia}, {Danese}, {Davies}, {Davis}, {de Bernardis}, {de Rosa}, {de
  Zotti}, {Delabrouille}, {Dickinson}, {Diego}, {Dole}, {Donzelli}, {Dor{\'e}},
  {Douspis}, {Ducout}, {Dupac}, {Efstathiou}, {Elsner}, {En{\ss}lin},
  {Eriksen}, {Falceta-Gon{\c{c}}alves}, {Falgarone}, {Ferri{\`e}re}, {Finelli},
  {Forni}, {Frailis}, {Fraisse}, {Franceschi}, {Frejsel}, {Galeotta}, {Galli},
  {Ganga}, {Ghosh}, {Giard}, {Gjerl{\o}w}, {Gonz{\'a}lez-Nuevo}, {G{\'o}rski},
  {Gregorio}, {Gruppuso}, {Gudmundsson}, {Guillet}, {Harrison}, {Helou},
  {Hennebelle}, {Henrot-Versill{\'e}}, {Hern{\'a}ndez-Monteagudo}, {Herranz},
  {Hildebrand t}, {Hivon}, {Holmes}, {Hornstrup}, {Huffenberger}, {Hurier},
  {Jaffe}, {Jaffe}, {Jones}, {Juvela}, {Keih{\"a}nen}, {Keskitalo}, {Kisner},
  {Knoche}, {Kunz}, {Kurki-Suonio}, {Lagache}, {Lamarre}, {Lasenby},
  {Lattanzi}, {Lawrence}, {Leonardi}, {Levrier}, {Liguori}, {Lilje},
  {Linden-V{\o}rnle}, {L{\'o}pez-Caniego}, {Lubin}, {Mac{\'\i}as-P{\'e}rez},
  {Maino}, {Mandolesi}, {Mangilli}, {Maris}, {Martin},
  {Mart{\'\i}nez-Gonz{\'a}lez}, {Masi}, {Matarrese}, {Melchiorri}, {Mendes},
  {Mennella}, {Migliaccio}, {Miville-Desch{\^e}nes}, {Moneti}, {Montier},
  {Morgante}, {Mortlock}, {Munshi}, {Murphy}, {Naselsky}, {Nati},
  {Netterfield}, {Noviello}, {Novikov}, {Novikov}, {Oppermann}, {Oxborrow},
  {Pagano}, {Pajot}, {Paladini}, {Paoletti}, {Pasian}, {Perotto}, {Pettorino},
  {Piacentini}, {Piat}, {Pierpaoli}, {Pietrobon}, {Plaszczynski},
  {Pointecouteau}, {Polenta}, {Ponthieu}, {Pratt}, {Prunet}, {Puget}, {Rachen},
  {Reinecke}, {Remazeilles}, {Renault}, {Renzi}, {Ristorcelli}, {Rocha},
  {Rossetti}, {Roudier}, {Rubi{\~n}o-Mart{\'\i}n}, {Rusholme}, {Sandri},
  {Santos}, {Savelainen}, {Savini}, {Scott}, {Soler}, {Stolyarov}, {Sudiwala},
  {Sutton}, {Suur-Uski}, {Sygnet}, {Tauber}, {Terenzi}, {Toffolatti}, {Tomasi},
  {Tristram}, {Tucci}, {Umana}, {Valenziano}, {Valiviita}, {Van Tent},
  {Vielva}, {Villa}, {Wade}, {Wandelt}, {Wehus}, {Ysard}, {Yvon}, \&
  {Zonca}}]{PlanckCollaborationetal2016}
---. 2016{\natexlab{a}}, \aap, 586, A138, \dodoi{10.1051/0004-6361/201525896}

\bibitem[{{Planck Collaboration} {et~al.}(2016{\natexlab{b}}){Planck
  Collaboration}, {Ade}, {Aghanim}, {Alves}, {Arnaud}, {Arzoumanian}, {Aumont},
  {Baccigalupi}, {Banday}, {Barreiro}, {Bartolo}, {Battaner}, {Benabed},
  {Benoit-L{\'e}vy}, {Bernard}, {Bern{\'e}}, {Bersanelli}, {Bielewicz},
  {Bonaldi}, {Bonavera}, {Bond}, {Borrill}, {Bouchet}, {Boulanger}, {Bracco},
  {Burigana}, {Calabrese}, {Cardoso}, {Catalano}, {Chamballu}, {Chiang},
  {Christensen}, {Clements}, {Colombi}, {Colombo}, {Combet}, {Couchot},
  {Crill}, {Curto}, {Cuttaia}, {Danese}, {Davies}, {Davis}, {de Bernardis}, {de
  Rosa}, {de Zotti}, {Delabrouille}, {Dickinson}, {Diego}, {Donzelli},
  {Dor{\'e}}, {Douspis}, {Ducout}, {Dupac}, {Elsner}, {En{\ss}lin}, {Eriksen},
  {Falgarone}, {Ferri{\`e}re}, {Finelli}, {Forni}, {Frailis}, {Fraisse},
  {Franceschi}, {Frejsel}, {Galeotta}, {Galli}, {Ganga}, {Ghosh}, {Giard},
  {Giraud-H{\'e}raud}, {Gjerl{\o}w}, {Gonz{\'a}lez-Nuevo}, {G{\'o}rski},
  {Gregorio}, {Gruppuso}, {Guillet}, {Hansen}, {Hanson}, {Harrison},
  {Hern{\'a}ndez-Monteagudo}, {Herranz}, {Hildebrandt}, {Hivon}, {Hobson},
  {Holmes}, {Huffenberger}, {Hurier}, {Jaffe}, {Jaffe}, {Jones}, {Juvela},
  {Keskitalo}, {Kisner}, {Knoche}, {Kunz}, {Kurki-Suonio}, {Lagache},
  {Lamarre}, {Lasenby}, {Lawrence}, {Leonardi}, {Levrier}, {Liguori}, {Lilje},
  {Linden-V{\o}rnle}, {L{\'o}pez-Caniego}, {Lubin}, {Mac{\'\i}as-P{\'e}rez},
  {Maffei}, {Mandolesi}, {Mangilli}, {Maris}, {Martin},
  {Mart{\'\i}nez-Gonz{\'a}lez}, {Masi}, {Matarrese}, {Mazzotta}, {Melchiorri},
  {Mendes}, {Mennella}, {Migliaccio}, {Mitra}, {Miville-Desch{\^e}nes},
  {Moneti}, {Montier}, {Morgante}, {Mortlock}, {Munshi}, {Murphy}, {Naselsky},
  {Nati}, {Natoli}, {N{\o}rgaard-Nielsen}, {Noviello}, {Novikov}, {Novikov},
  {Oppermann}, {Pagano}, {Pajot}, {Paladini}, {Paoletti}, {Pasian}, {Perrotta},
  {Pettorino}, {Piacentini}, {Piat}, {Pierpaoli}, {Pietrobon}, {Plaszczynski},
  {Pointecouteau}, {Polenta}, {Pratt}, {Puget}, {Rachen}, {Rebolo}, {Reinecke},
  {Remazeilles}, {Renault}, {Renzi}, {Ricciardi}, {Ristorcelli}, {Rocha},
  {Rosset}, {Rossetti}, {Roudier}, {Rubi{\~n}o-Mart{\'\i}n}, {Rusholme},
  {Sandri}, {Savelainen}, {Savini}, {Scott}, {Soler}, {Stolyarov}, {Sutton},
  {Suur-Uski}, {Sygnet}, {Tauber}, {Terenzi}, {Toffolatti}, {Tomasi},
  {Tristram}, {Tucci}, {Valenziano}, {Valiviita}, {Van Tent}, {Vielva},
  {Villa}, {Wade}, {Wandelt}, {Yvon}, {Zacchei}, \&
  {Zonca}}]{PlanckCollaborationXXXIII2016a}
---. 2016{\natexlab{b}}, \aap, 586, A136, \dodoi{10.1051/0004-6361/201425305}

\bibitem[{{Punanova} {et~al.}(2018){Punanova}, {Caselli}, {Pineda}, {Pon},
  {Tafalla}, {Hacar}, \& {Bizzocchi}}]{Punanovaetal2018}
{Punanova}, A., {Caselli}, P., {Pineda}, J.~E., {et~al.} 2018, \aap, 617, A27,
  \dodoi{10.1051/0004-6361/201731159}

\bibitem[{{Roy} {et~al.}(2014){Roy}, {Andr{\'e}}, {Palmeirim}, {Attard},
  {K{\"o}nyves}, {Schneider}, {Peretto}, {Men'shchikov}, {Ward-Thompson},
  {Kirk}, {Griffin}, {Marsh}, {Abergel}, {Arzoumanian}, {Benedettini}, {Hill},
  {Motte}, {Nguyen Luong}, {Pezzuto}, {Rivera-Ingraham}, {Roussel}, {Rygl},
  {Spinoglio}, {Stamatellos}, \& {White}}]{Royetal2014}
{Roy}, A., {Andr{\'e}}, P., {Palmeirim}, P., {et~al.} 2014, \aap, 562, A138,
  \dodoi{10.1051/0004-6361/201322236}

\bibitem[{{Shimajiri} {et~al.}(2019){Shimajiri}, {Andr{\'e}}, {Palmeirim},
  {Arzoumanian}, {Bracco}, {K{\"o}nyves}, {Ntormousi}, \&
  {Ladjelate}}]{Shimajirietal2019}
{Shimajiri}, Y., {Andr{\'e}}, P., {Palmeirim}, P., {et~al.} 2019, \aap, 623,
  A16, \dodoi{10.1051/0004-6361/201834399}

\bibitem[{{Shu} {et~al.}(1987){Shu}, {Adams}, \& {Lizano}}]{Shuetal1987}
{Shu}, F.~H., {Adams}, F.~C., \& {Lizano}, S. 1987, \araa, 25, 23,
  \dodoi{10.1146/annurev.aa.25.090187.000323}

\bibitem[{{Soam} {et~al.}(2018){Soam}, {Pattle}, {Ward-Thompson}, {Lee},
  {Sadavoy}, {Koch}, {Kim}, {Kwon}, {Kwon}, {Arzoumanian}, {Berry}, {Hoang},
  {Tamura}, {Lee}, {Liu}, {Kim}, {Johnstone}, {Nakamura}, {Lyo}, {Onaka},
  {Kim}, {Furuya}, {Hasegawa}, {Lai}, {Bastien}, {Chung}, {Kim}, {Parsons},
  {Rawlings}, {Mairs}, {Graves}, {Robitaille}, {Liu}, {Whitworth}, {Eswaraiah},
  {Rao}, {Yoo}, {Houde}, {Kang}, {Doi}, {Choi}, {Kang}, {Coud{\'e}}, {Li},
  {Matsumura}, {Matthews}, {Pon}, {Di Francesco}, {Hayashi}, {Kawabata},
  {Inutsuka}, {Qiu}, {Franzmann}, {Friberg}, {Greaves}, {Kirk}, {Li},
  {Shinnaga}, {van Loo}, {Aso}, {Byun}, {Chen}, {Chen}, {Chen}, {Ching}, {Cho},
  {Chrysostomou}, {Drabek-Maunder}, {Eyres}, {Fiege}, {Friesen}, {Fuller},
  {Gledhill}, {Griffin}, {Gu}, {Hatchell}, {Holland }, {Inoue}, {Iwasaki},
  {Jeong}, {Kang}, {Kemper}, {Kim}, {Kim}, {Lacaille}, {Lee}, {Li}, {Liu},
  {Liu}, {Moriarty-Schieven}, {Nakanishi}, {Ohashi}, {Peretto}, {Pyo}, {Qian},
  {Retter}, {Richer}, {Rigby}, {Savini}, {Scaife}, {Tang}, {Tomisaka}, {Wang},
  {Wang}, {Yen}, {Yuan}, {Zhang}, {Zhang}, {Zhou}, {Zhu}, {Andr{\'e}},
  {Dowell}, {Falle}, {Tsukamoto}, {Kanamori}, {Kataoka}, {Kobayashi}, {Nagata},
  {Saito}, {Seta}, {Hwang}, {Han}, {Lee}, \& {Zenko}}]{Soametal2018}
{Soam}, A., {Pattle}, K., {Ward-Thompson}, D., {et~al.} 2018, \apj, 861, 65,
  \dodoi{10.3847/1538-4357/aac4a6}

\bibitem[{{Sokolov} {et~al.}(2018){Sokolov}, {Wang}, {Pineda}, {Caselli},
  {Henshaw}, {Barnes}, {Tan}, {Fontani}, {Jim{\'e}nez-Serra}, \&
  {Zhang}}]{Sokolovetal2018}
{Sokolov}, V., {Wang}, K., {Pineda}, J.~E., {et~al.} 2018, \aap, 611, L3,
  \dodoi{10.1051/0004-6361/201832746}

\bibitem[{{Soler} \& {Hennebelle}(2017)}]{Soleretal2017}
{Soler}, J.~D., \& {Hennebelle}, P. 2017, \aap, 607, A2,
  \dodoi{10.1051/0004-6361/201731049}

\bibitem[{{Soler} {et~al.}(2016){Soler}, {Alves}, {Boulanger}, {Bracco},
  {Falgarone}, {Franco}, {Guillet}, {Hennebelle}, {Levrier}, {Martin}, \&
  {Miville-Desch{\^e}nes}}]{Soleretal2016}
{Soler}, J.~D., {Alves}, F., {Boulanger}, F., {et~al.} 2016, \aap, 596, A93,
  \dodoi{10.1051/0004-6361/201628996}

\bibitem[{{Stephens} {et~al.}(2013){Stephens}, {Looney}, {Kwon}, {Hull},
  {Plambeck}, {Crutcher}, {Chapman}, {Novak}, {Davidson}, {Vaillancourt},
  {Shinnaga}, \& {Matthews}}]{Stephensetal2013}
{Stephens}, I.~W., {Looney}, L.~W., {Kwon}, W., {et~al.} 2013, \apjl, 769, L15,
  \dodoi{10.1088/2041-8205/769/1/L15}

\bibitem[{{Stone} {et~al.}(1998){Stone}, {Ostriker}, \&
  {Gammie}}]{Stoneetal1998}
{Stone}, J.~M., {Ostriker}, E.~C., \& {Gammie}, C.~F. 1998, \apjl, 508, L99,
  \dodoi{10.1086/311718}

\bibitem[{{Sugitani} {et~al.}(2010){Sugitani}, {Nakamura}, {Tamura},
  {Watanabe}, {Kandori}, {Nishiyama}, {Kusakabe}, {Hashimoto}, {Nagata}, \&
  {Sato}}]{Sugitanietal2010}
{Sugitani}, K., {Nakamura}, F., {Tamura}, M., {et~al.} 2010, \apj, 716, 299,
  \dodoi{10.1088/0004-637X/716/1/299}

\bibitem[{{Tafalla} \& {Hacar}(2015)}]{TafallaHacar2015}
{Tafalla}, M., \& {Hacar}, A. 2015, \aap, 574, A104,
  \dodoi{10.1051/0004-6361/201424576}

\bibitem[{{Tafalla} {et~al.}(2010){Tafalla}, {Santiago-Garc{\'\i}a}, {Hacar},
  \& {Bachiller}}]{Tafallaetal2010}
{Tafalla}, M., {Santiago-Garc{\'\i}a}, J., {Hacar}, A., \& {Bachiller}, R.
  2010, \aap, 522, A91, \dodoi{10.1051/0004-6361/201015158}

\bibitem[{{Takakuwa} {et~al.}(2018){Takakuwa}, {Tsukamoto}, {Saigo}, \&
  {Saito}}]{Takakuwaetal2018}
{Takakuwa}, S., {Tsukamoto}, Y., {Saigo}, K., \& {Saito}, M. 2018, \apj, 865,
  51, \dodoi{10.3847/1538-4357/aadb93}

\bibitem[{{Tokuda} {et~al.}(2020){Tokuda}, {Fujishiro}, {Tachihara},
  {Takashima}, {Fukui}, {Zahorecz}, {Saigo}, {Matsumoto}, {Tomida}, {Machida},
  {Inutsuka}, {Andr{\'e}}, {Kawamura}, \& {Onishi}}]{Tokudaetal2020}
{Tokuda}, K., {Fujishiro}, K., {Tachihara}, K., {et~al.} 2020, \apj, 899, 10,
  \dodoi{10.3847/1538-4357/ab9ca7}

\bibitem[{{Wang} {et~al.}(2020){Wang}, {Lai}, {Clemens}, {Koch}, {Eswaraiah},
  {Chen}, \& {Pand ey}}]{WangJWetal2020}
{Wang}, J.-W., {Lai}, S.-P., {Clemens}, D.~P., {et~al.} 2020, \apj, 888, 13,
  \dodoi{10.3847/1538-4357/ab5c1c}

\bibitem[{{Wang} {et~al.}(2019){Wang}, {Lai}, {Eswaraiah}, {Pattle}, {Di
  Francesco}, {Johnstone}, {Koch}, {Liu}, {Tamura}, {Furuya}, {Onaka},
  {Ward-Thompson}, {Soam}, {Kim}, {Lee}, {Lee}, {Mairs}, {Arzoumanian}, {Kim},
  {Hoang}, {Hwang}, {Liu}, {Berry}, {Bastien}, {Hasegawa}, {Kwon}, {Qiu},
  {Andr{\'e}}, {Aso}, {Byun}, {Chen}, {Chen}, {Chen}, {Ching}, {Cho}, {Choi},
  {Chrysostomou}, {Chung}, {Coud{\'e}}, {Doi}, {Dowell}, {Drabek-Maunder},
  {Duan}, {Eyres}, {Falle}, {Fanciullo}, {Fiege}, {Franzmann}, {Friberg},
  {Friesen}, {Fuller}, {Gledhill}, {Graves}, {Greaves}, {Griffin}, {Gu}, {Han},
  {Hatchell}, {Hayashi}, {Holland}, {Houde}, {Inoue}, {Inutsuka}, {Iwasaki},
  {Jeong}, {Kanamori}, {Kang}, {Kang}, {Kang}, {Kataoka}, {Kawabata}, {Kemper},
  {Kim}, {Kim}, {Kim}, {Kim}, {Kirk}, {Kobayashi}, {Konyves}, {Kwon},
  {Lacaille}, {Lee}, {Lee}, {Lee}, {Lee}, {Li}, {Li}, {Li}, {Liu}, {Liu},
  {Lyo}, {Matsumura}, {Matthews}, {Moriarty-Schieven}, {Nagata}, {Nakamura},
  {Nakanishi}, {Ohashi}, {Park}, {Parsons}, {Pascale}, {Peretto}, {Pon}, {Pyo},
  {Qian}, {Rao}, {Rawlings}, {Retter}, {Richer}, {Rigby}, {Robitaille},
  {Sadavoy}, {Saito}, {Savini}, {Scaife}, {Seta}, {Shinnaga}, {Tang},
  {Tomisaka}, {Tsukamoto}, {van Loo}, {Wang}, {Whitworth}, {Yen}, {Yoo},
  {Yuan}, {Yun}, {Zenko}, {Zhang}, {Zhang}, {Zhang}, {Zhou}, \&
  {Zhu}}]{WangJWetal2019}
{Wang}, J.-W., {Lai}, S.-P., {Eswaraiah}, C., {et~al.} 2019, \apj, 876, 42,
  \dodoi{10.3847/1538-4357/ab13a2}

\bibitem[{{Wang} {et~al.}(2014){Wang}, {Shang}, {Su}, {Santiago-Garc{\'\i}a},
  {Tafalla}, {Zhang}, {Hirano}, \& {Lee}}]{WangLYetal2014}
{Wang}, L.-Y., {Shang}, H., {Su}, Y.-N., {et~al.} 2014, \apj, 780, 49,
  \dodoi{10.1088/0004-637X/780/1/49}

\bibitem[{{Ward-Thompson} {et~al.}(2020){Ward-Thompson}, {McKee}, {Furuya}, \&
  {Tsukamoto}}]{Ward-Thompsonetal2020}
{Ward-Thompson}, D., {McKee}, C.~F., {Furuya}, R., \& {Tsukamoto}, Y. 2020,
  Frontiers in Astronomy and Space Sciences, 7, 13,
  \dodoi{10.3389/fspas.2020.00013}

\bibitem[{{Ward-Thompson} {et~al.}(2017){Ward-Thompson}, {Pattle}, {Bastien},
  {Furuya}, {Kwon}, {Lai}, {Qiu}, {Berry}, {Choi}, {Coud{\'e}}, {Di Francesco},
  {Hoang}, {Franzmann}, {Friberg}, {Graves}, {Greaves}, {Houde}, {Johnstone},
  {Kirk}, {Koch}, {Kwon}, {Lee}, {Li}, {Matthews}, {Mottram}, {Parsons}, {Pon},
  {Rao}, {Rawlings}, {Shinnaga}, {Sadavoy}, {van Loo}, {Aso}, {Byun},
  {Eswaraiah}, {Chen}, {Chen}, {Chen}, {Ching}, {Cho}, {Chrysostomou}, {Chung},
  {Doi}, {Drabek-Maunder}, {Eyres}, {Fiege}, {Friesen}, {Fuller}, {Gledhill},
  {Griffin}, {Gu}, {Hasegawa}, {Hatchell}, {Hayashi}, {Holland}, {Inoue},
  {Inutsuka}, {Iwasaki}, {Jeong}, {Kang}, {Kang}, {Kang}, {Kawabata}, {Kemper},
  {Kim}, {Kim}, {Kim}, {Kim}, {Kim}, {Kim}, {Lacaille}, {Lee}, {Lee}, {Li},
  {Li}, {Liu}, {Liu}, {Liu}, {Liu}, {Lyo}, {Mairs}, {Matsumura},
  {Moriarty-Schieven}, {Nakamura}, {Nakanishi}, {Ohashi}, {Onaka}, {Peretto},
  {Pyo}, {Qian}, {Retter}, {Richer}, {Rigby}, {Robitaille}, {Savini}, {Scaife},
  {Soam}, {Tamura}, {Tang}, {Tomisaka}, {Wang}, {Wang}, {Whitworth}, {Yen},
  {Yoo}, {Yuan}, {Zhang}, {Zhang}, {Zhou}, {Zhu}, {Andr{\'e}}, {Dowell},
  {Falle}, \& {Tsukamoto}}]{Ward-Thompsonetal2017}
{Ward-Thompson}, D., {Pattle}, K., {Bastien}, P., {et~al.} 2017, \apj, 842, 66,
  \dodoi{10.3847/1538-4357/aa70a0}

\bibitem[{{Wurster} \& {Lewis}(2020)}]{WursterLewis2020}
{Wurster}, J., \& {Lewis}, B.~T. 2020, \mnras, 495, 3795,
  \dodoi{10.1093/mnras/staa1339}

\bibitem[{{Xu} {et~al.}(2020){Xu}, {Li}, {Dai}, {Goldsmith}, \&
  {Fuller}}]{XuXetal2020}
{Xu}, X., {Li}, D., {Dai}, Y.~S., {Goldsmith}, P.~F., \& {Fuller}, G.~A. 2020,
  \apj, 898, 122, \dodoi{10.3847/1538-4357/ab9a45}

\bibitem[{{Yen} {et~al.}(2020){Yen}, {Koch}, {Hull}, {Ward-Thompson},
  {Bastien}, {Hasegawa}, {Kwon}, {Lai}, {Qiu}, {Ching}, {Chung}, {Coude}, {Di
  Francesco}, {Diep}, {Doi}, {Eswaraiah}, {Falle}, {Fuller}, {Furuya}, {Han},
  {Hatchell}, {Houde}, {Inutsuka}, {Johnstone}, {Kang}, {Kang}, {Kim},
  {Kirchschlager}, {Kwon}, {Lee}, {Lee}, {Liu}, {Liu}, {Lyo}, {Ohashi},
  {Onaka}, {Pattle}, {Sadavoy}, {Saito}, {Shinnaga}, {Soam}, {Tahani},
  {Tamura}, {Tang}, {Tang}, \& {Zhang}}]{Yenetal2020}
{Yen}, H.-W., {Koch}, P.~M., {Hull}, C. L.~H., {et~al.} 2020, arXiv e-prints,
  arXiv:2011.06731.
\newblock \doarXiv{2011.06731}

\end{thebibliography}

\begin{table*}
\centering
\caption{Polarization data along with the celestial coordinates of the pixels.}
\label{tab:poldata}
\begin{tabular}{cccccccc} %
\hline
\hline

\hline
RA (J2000) &  Dec (J2000) & $I \pm \sigma_{I}$  &  $Q \pm \sigma_{Q}$ & $U \pm \sigma_{U}$ & $PI \pm \sigma_{PI}$ &  $P \pm \sigma_{P}$  & $\theta_{\mathrm{core, B}} \pm \sigma_{\theta_{\mathrm{core,B}}}$  \\
($\degr$)   &   ($\degr$)     &  (mJy beam$^{-1}$) & (mJy beam$^{-1}$) & (mJy beam$^{-1}$) & (mJy beam$^{-1}$) & (\%) &   ($\degr$)  \\
(1)        &         (2)        &   (3)  &     (4)     &   (5)            &   (6)               &  (7)  &  (8)    \\
                   \hline
64.982471 & 27.157917  &  25.0 $\pm$ 0.6  & -0.7 $\pm$ 0.6 & -2.5 $\pm$  0.7 & 2.5 $\pm$ 0.7  &   9.8 $\pm$ 2.8  &  37 $\pm$  7   \\
65.004946 & 27.161250  &  16.9 $\pm$ 0.7  &  1.1 $\pm$ 0.7 &  2.7 $\pm$  0.7 & 2.8 $\pm$ 0.7  &  16.8 $\pm$ 4.4  & 124 $\pm$  7   \\
64.989963 & 27.161250  &  33.0 $\pm$ 0.7  &  0.3 $\pm$ 0.6 &  3.0 $\pm$  0.6 & 3.0 $\pm$ 0.6  &   9.0 $\pm$ 1.9  & 132 $\pm$  6   \\
64.986217 & 27.161250  &  37.3 $\pm$ 0.7  & -0.7 $\pm$ 0.6 &  3.1 $\pm$  0.7 & 3.1 $\pm$ 0.7  &   8.3 $\pm$ 1.8  & 141 $\pm$  5   \\
65.004946 & 27.164583  &  35.7 $\pm$ 0.7  &  1.1 $\pm$ 0.7 &  2.7 $\pm$  0.7 & 2.9 $\pm$ 0.7  &   8.1 $\pm$ 1.9  & 124 $\pm$  7   \\
65.001200 & 27.164583  &  78.6 $\pm$ 0.7  & -0.2 $\pm$ 0.7 &  2.7 $\pm$  0.7 & 2.6 $\pm$ 0.7  &   3.3 $\pm$ 0.9  & 137 $\pm$  7   \\
64.989963 & 27.164583  & 113.9 $\pm$ 0.9  &  2.3 $\pm$ 0.6 &  0.3 $\pm$  0.6 & 2.2 $\pm$ 0.6  &   1.9 $\pm$ 0.5  &  94 $\pm$  8   \\
65.008696 & 27.167917  &  20.3 $\pm$ 0.6  & -1.6 $\pm$ 0.7 &  1.6 $\pm$  0.7 & 2.2 $\pm$ 0.7  &  10.8 $\pm$ 3.3  & 157 $\pm$  8   \\
64.993708 & 27.167917  & 318.4 $\pm$ 1.7  &  2.7 $\pm$ 0.7 &  0.2 $\pm$  0.6 & 2.6 $\pm$ 0.7  &   0.8 $\pm$ 0.2  &  92 $\pm$  7   \\
64.989963 & 27.167917  & 140.9 $\pm$ 0.8  &  3.1 $\pm$ 0.6 &  1.6 $\pm$  0.6 & 3.5 $\pm$ 0.6  &   2.5 $\pm$ 0.4  & 104 $\pm$  5   \\
64.986212 & 27.174583  &  45.4 $\pm$ 0.6  &  1.9 $\pm$ 0.6 &  0.9 $\pm$  0.6 & 2.0 $\pm$ 0.6  &   4.4 $\pm$ 1.3  & 103 $\pm$  8   \\
64.967479 & 27.181247  &  12.2 $\pm$ 0.5  & -1.5 $\pm$ 0.6 & -1.6 $\pm$  0.6 & 2.1 $\pm$ 0.6  &  17.6 $\pm$ 5.1  &  24 $\pm$  8   \\
64.971225 & 27.184581  &  15.8 $\pm$ 0.6  &  1.4 $\pm$ 0.6 & -1.9 $\pm$  0.7 & 2.3 $\pm$ 0.6  &  14.4 $\pm$ 4.0  &  63 $\pm$  7   \\
64.959979 & 27.191247  &  46.1 $\pm$ 0.6  &  0.2 $\pm$ 0.6 &  2.0 $\pm$  0.6 & 1.9 $\pm$ 0.6  &   4.1 $\pm$ 1.3  & 132 $\pm$  9   \\
64.971221 & 27.194583  &  26.1 $\pm$ 0.6  &  0.2 $\pm$ 0.6 &  2.9 $\pm$  0.6 & 2.9 $\pm$ 0.6  &  11.0 $\pm$ 2.5  & 133 $\pm$  6   \\
64.967475 & 27.194581  &  48.4 $\pm$ 0.6  & -2.0 $\pm$ 0.6 &  0.4 $\pm$  0.6 & 1.9 $\pm$ 0.6  &   4.0 $\pm$ 1.3  & 174 $\pm$  9   \\
64.963729 & 27.194581  &  58.6 $\pm$ 0.6  & -1.5 $\pm$ 0.6 & -1.8 $\pm$  0.6 & 2.3 $\pm$ 0.6  &   3.8 $\pm$ 1.0  &  25 $\pm$  7   \\
64.933733 & 27.217906  &  54.7 $\pm$ 0.6  & -0.8 $\pm$ 0.6 & -2.4 $\pm$  0.6 & 2.5 $\pm$ 0.6  &   4.6 $\pm$ 1.1  &  36 $\pm$  6   \\
64.926237 & 27.217903  &  25.0 $\pm$ 0.6  & -0.6 $\pm$ 0.5 & -2.7 $\pm$  0.7 & 2.7 $\pm$ 0.7  &  10.8 $\pm$ 2.7  &  39 $\pm$  6   \\
64.926237 & 27.221236  &  72.0 $\pm$ 0.6  &  0.4 $\pm$ 0.6 & -2.9 $\pm$  0.6 & 2.8 $\pm$ 0.6  &   3.9 $\pm$ 0.8  &  49 $\pm$  6   \\
64.937479 & 27.227906  &  37.7 $\pm$ 0.6  &  1.8 $\pm$ 0.6 & -1.9 $\pm$  0.7 & 2.5 $\pm$ 0.6  &   6.7 $\pm$ 1.7  &  66 $\pm$  7   \\
64.933729 & 27.227906  & 113.4 $\pm$ 0.7  &  2.2 $\pm$ 0.6 & -0.0 $\pm$  0.7 & 2.1 $\pm$ 0.6  &   1.8 $\pm$ 0.6  &  90 $\pm$  9   \\
64.929979 & 27.227903  & 290.0 $\pm$ 1.4  &  1.2 $\pm$ 0.6 & -2.3 $\pm$  0.6 & 2.5 $\pm$ 0.6  &   0.9 $\pm$ 0.2  &  59 $\pm$  7   \\
64.926233 & 27.227903  & 253.9 $\pm$ 1.5  & -1.8 $\pm$ 0.6 & -2.2 $\pm$  0.7 & 2.8 $\pm$ 0.6  &   1.1 $\pm$ 0.3  &  25 $\pm$  6   \\
64.933725 & 27.234572  &  14.6 $\pm$ 0.6  &  0.5 $\pm$ 0.7 & -2.2 $\pm$  0.7 & 2.2 $\pm$ 0.7  &  14.9 $\pm$ 4.5  &  51 $\pm$  9   \\
64.854983 & 27.247850  &  29.3 $\pm$ 1.1  &  3.6 $\pm$ 1.0 & -2.0 $\pm$  1.0 & 3.9 $\pm$ 1.0  &  13.4 $\pm$ 3.5  &  76 $\pm$  7   \\
64.907471 & 27.251225  &  22.1 $\pm$ 0.8  &  4.0 $\pm$ 0.8 & -0.5 $\pm$  0.8 & 3.9 $\pm$ 0.8  &  17.8 $\pm$ 3.8  &  86 $\pm$  6   \\
64.922458 & 27.267900  &  91.9 $\pm$ 1.3  &  1.3 $\pm$ 0.9 &  3.7 $\pm$  0.9 & 3.8 $\pm$ 0.9  &   4.2 $\pm$ 1.0  & 125 $\pm$  7   \\
        \hline
\end{tabular} \\
\tablecomments{RA and Dec: Celestial coordinates.}
\tablecomments{$I$: Total intensity.}
\tablecomments{$Q$ and $U$: Stokes parameters.}
\tablecomments{$PI$: Debiased polarized intensity.}
\tablecomments{$P$: Debiased degree of polarization}
\tablecomments{$\theta_{\mathrm{core,B}}$: B-field orientation, determined by applying an offset of 90$\degr$ to the polarization angle.}
\end{table*}

\begin{table*}[!ht]
\centering
\scriptsize
        \caption{Various parameters for K04166, K04169, and Miz-8b.}
        \label{tab:paramscl12}
        \begin{tabular}{clccc}\hline 
		No &    Parameter &  K04166 &  K04169 & Miz-8b \\
		\hline
		& Weighted mean B-field orientation along with various offset position angles. & & & \\
		\hline
 1 & No. of B-field segments & 	8   & 10 &  4 \\
2 &  Weighted mean B-field orientation ($\bar{\theta}_{\mathrm{core, B}}$~$\pm\epsilon_{\mathrm{\bar{\theta}_{\mathrm{core, B}}}}; \degr$)$^{a,b}$ & 48 $\pm$ 2 & 121 $\pm$ 2 &  158 $\pm$ 4  \\
3 & Angular dispersion ($\delta_{\theta}$, $\degr$) &  18$\pm$4
 &  20$\pm$3 & 35$\pm$7 \\
4  & Standard deviation in ${\theta_{\mathrm{core, B}}}~~($\degr$)^{b}$ & 20 & 33 & 35 \\
5  & Position angle of core major axis ($\theta_{\mathrm{core}}$; $\degr$)$^{c}$ & 127$\pm$20 & 126$\pm$16 & 119$\pm$30 \\
6  & $|\bar{\theta}_{\mathrm{core, B}}-\theta_{\mathrm{B}}^{\mathrm{largescale}}|$ ($\degr$)$^{d}$ &
        19 & 88 & 51 \\
7  & $|\bar{\theta}_{\mathrm{core, B}}-\theta_{\mathrm{core}}|$ ($\degr$) & 79 & 5 & 39 \\
       \hline
 & Core dimensions, mass, and column and number densities, etc. & & & \\
 \hline
 1 & Semi-major axis (a) (pc) & 0.032$\pm$0.005 & 0.036$\pm$0.006 &  0.029 $\pm$0.010 \\
2 & Semi-minor axis (b) (pc) & 0.024$\pm$0.003 & 0.025$\pm$0.005 & 0.023   $\pm$  0.005 \\
3 & Effective radius R$_{\mathrm{eff}}$ (pc)$^{e}$  & 0.028 $\pm$   0.003 & 0.030 $\pm$   0.004 &   0.026 $\pm$ 0.005 \\
4 & Median dust temperature (T$_{\mathrm{d}}$)  (K)$^{f}$  & 12.2 $\pm$  0.4  &  12.2 $\pm$  0.8 & 10.9 $\pm$ 0.2 \\
5 & Integrated flux ($F_{\mathrm {\nu}}$)  (mJy) & 1564  & 1969 &  516 \\
6 & Mass (M) (M$_{\sun}$) &  0.55$\pm$0.28 & 0.69$\pm$0.36 &   0.22 $\pm$ 0.11   \\
7 & Column density ($N(\mathrm{H_{2}})$) $\times$10$^{21}$ (cm$^{-2}$) & 10 $\pm$ 6 & 11 $\pm$ 6 &  5$\pm$3 \\
8 & Number density ($n(\mathrm{H_{2}})$) $\times$10$^{4}$ (cm$^{-3}$)  & 9 $\pm$ 5  & 9 $\pm$ 6 &   5$\pm$4 \\
\hline
 & B-field strength, and magnetic and turbulent pressures, etc. & & \\
\hline
1 & B-field strength (using DCF method) ($\mu$G)  & 38$\pm$14   & 44$\pm$16  & 12$\pm$5 \\
2 & B-field pressure ($P_{\mathrm{B}}$; $\times$10$^{-10}$ dyn cm$^{-2}$)   & 0.6$\pm$0.4   & 0.8$\pm$0.6 & 0.06$\pm$0.05 \\
3 & Turbulent pressure ($P_{\mathrm{turb}}$; $\times$10$^{-10}$ dyn cm$^{-2}$) & 0.4$\pm$0.3 & 0.7$\pm$0.5 & 0.2$\pm$0.1 \\
4 & $P_{\mathrm{B}}/P_{\mathrm{turb}}$    & 1.3$\pm$1.3     & 1.0$\pm$1.0 &  0.3$\pm$0.4
  \\
5 & Mass-to-flux ratio criticality ($\lambda$~$=$~$(M/\phi)$/$(M/\phi)_{\mathrm{cri}}$) & 0.7$\pm$0.4 & 1.4$\pm$1.0 & 3$\pm$2 \\
6 & Alfv{\'e}n velocity ($V_{\mathrm{A}}$; km s$^{-1}$) & 0.17$\pm$0.09 & 0.21$\pm$0.12 &   0.07$\pm$0.04  \\
7 & Alfv{\'e}nic Mach number ($M_{\mathrm{A}}$) & 1.1$\pm$0.6 & 1.2$\pm$0.7 & 2$\pm$1\\
\hline
 & Energy parameters & & \\
 \hline
1 & $p$  & 0.27 & 0.27 & 0.27 \\
2  & Rotational energy $E_{\mathrm{rot}}$ ($\times10^{41}$ erg) & 0.02 & 0.1 & 0.01   \\ 
3 & Magnetic energy $E_{\mathrm{mag}}$ ($\times10^{41}$ erg) &   2   & 3  & 0.1  \\
4 & $E_{\mathrm{rot}}/E_{\mathrm{mag}}$ & 0.01 & 0.04 & 0.05  \\
 \hline
  & Various position angles (PAs) & & \\
\hline
1 & Core minor axis PA ($\degr$) &  $\sim$37       &    $\sim$36 & $\sim$29  \\
2 & Outflows PA ($\degr$)$^{g}$      &    $\sim$33       &    $\sim$58 & --  \\
3 & Core ${\theta_{\mathrm{G}}}$ ($\degr$; from W to N)$^{h}$  &  $-$169  & $-$144 & 93 \\
4 & Core $\theta_{\mathrm{G}}$ ($\degr$; from N to E)$^{i}$ & 101 & 126 & 3 \\
5 & PA of the core rotation-axis ($\degr$; from N to E)$^{j}$ &  11 & 36 & 93 \\
\hline
\end{tabular}\\
\tablecomments{$a$ While estimating $\bar{\theta}_{\mathrm{core, B}}$, one B-field segment associated with K04169, with PA $\sim37\degr$, was ignored as it belongs to an another condensation to the south of K04169. Similarly, two segments associated with Miz-8b, with PAs of 24$\degr$ and 63$\degr$, were ignored as they fall in the core boundary and do not represent the B-field orientation in Miz-8b. These ignored segments are, within the 30$\degr$, parallel to the large-scale B-field (with PA of 29\degr)}. 
\tablecomments{$b$: We also cross-check the $\bar{\theta}_{\mathrm{core, B}}$ and corresponding standard deviation values against the circular mean and circular standard deviation values \citep[][see their Appendix C]{Doietal2020}, which are estimated to be 51$\pm$19$\degr$, 121$\pm$21$\degr$, and 158$\pm$35$\degr$ for K04166, K04169, and Miz-8b, respectively. These values are quite consistent with the quoted weighted mean and standard deviations.}
		   \tablecomments{$c$: $\theta_{\mathrm{core}}$ is the position angle of the major axis of the core obtained by fitting an ellipse to the 13 mJy beam$^{-1}$ POL-2 Stokes I contours of the cores.}
\tablecomments{$d$: $\theta_{B}^{\mathrm{largescale}}$ is the mean large scale mean B-field orientation (29$\degr$;
        Table \ref{tab:cloudpas}).}
	\tablecomments{$e$: $R_{\mathrm{eff}}$~$=$~$\sqrt{ab}$.}
	\tablecomments{$f$: Based on the {\it Herschel} Gould Belt Survey  (HGBS) column density map.}
	\tablecomments{$g$: The PA of outflows is determined based on the midline that passes through the center of the bipolar cones.  The outflow data is from \citep[][see also \citealt{Ohashietal1997}]{Tokudaetal2020}.}
	\tablecomments{$h$: Core $\theta_{G}$ is based on the total velocity gradient derived from N$_{2}$H$^{+}$ data \citep{Punanovaetal2018}. Negative sign corresponds to the angle in clock-wise direction from W to S (see Table B.1 of \citealt{Punanovaetal2018}).}
\tablecomments{${i, j}$: Core $\theta_{G}$ and PA of the core rotation are perpendicular to each other.}
\end{table*}

\appendix

\restartappendixnumbering

\section{Polarization properties: detection of weakly polarized dust emission}\label{subsec:weakpi}

\begin{figure*}
 \centering
	 \resizebox{8.5cm}{20cm}{\includegraphics{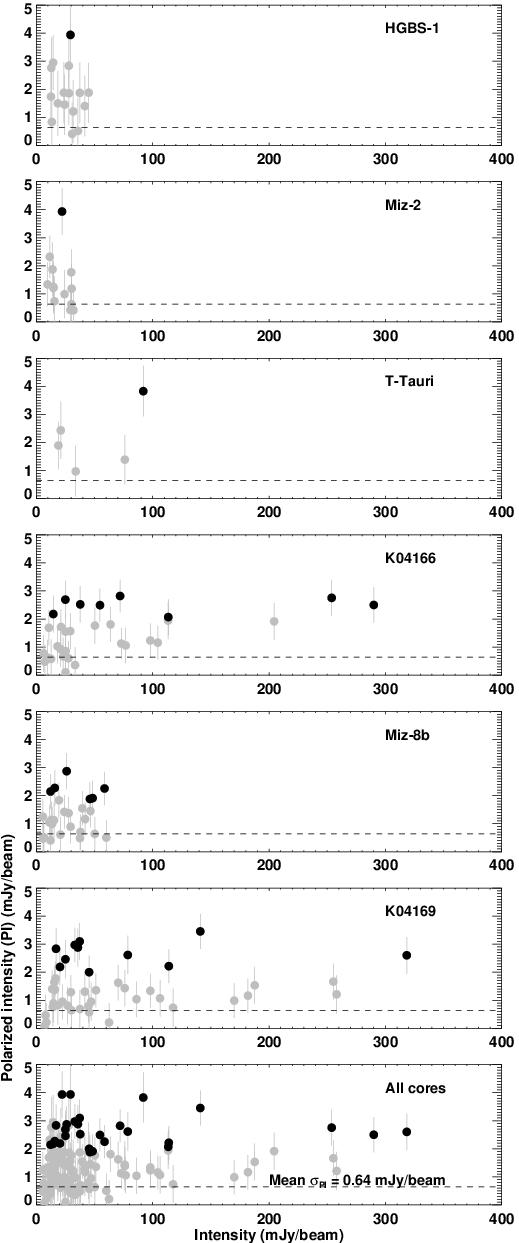}}
	 \caption{Polarized intensity ($PI$) versus intensity ($I$) plots for each core, and all of the cores combined. The name of the core is stated in each panel. Gray filled circles represent the data satisfying the criterion $I/\sigma_{I}$~$>$~10, whereas the black filled circles denote those satisfying both the criteria $I/\sigma_{I}$~$>$~10 and $P/\sigma_{P}$~$\geq$~3.
         The dashed line represents the median $\sigma_{PI}$~$=$~0.64 mJy/beam determined from the filled black points. }\label{fig:ivspi}
\end{figure*}

Figure \ref{fig:ivspi} plots $PI$ versus $I$ for each core, using the selection criterion $I/\sigma_{I}$~$>$~10 (gray
filled circles). In at least three cores, K04166, Miz-8b, and K04169, and also in the plot showing all of the cores,
a slowly increasing trend in $PI$ can be seen up to I $\sim$100 mJy beam$^{-1}$, beyond which $PI$ remains approximately constant, although there exist fewer data points. 

\begin{figure*}
\centering
	\resizebox{13cm}{12cm}{\includegraphics{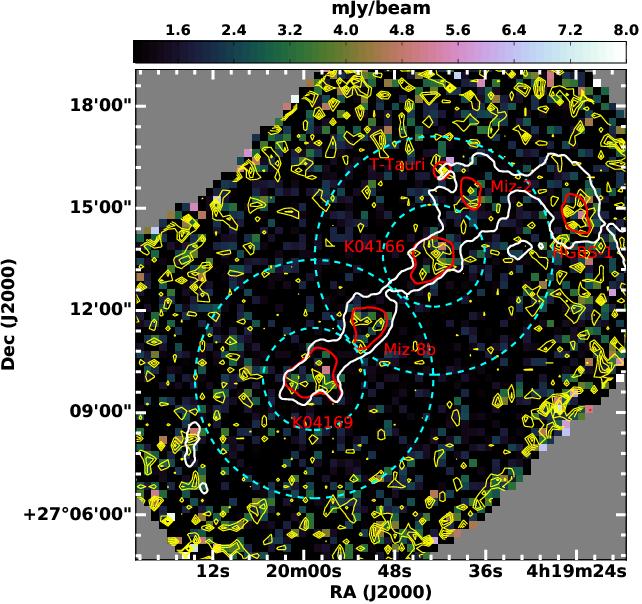}}
	\caption{Debiased polarized intensity (PI) map produced using our POL-2 Stokes $Q$ and $U$ maps of the B213 region. 
	Non-smoothed PI contours are drawn at [2, 3, 4]$\times\sigma_{PI}$, where $\sigma_{PI}$ is the rms noise, $\sim$1 mJy beam$^{-1}$ (estimated using the pixels in a signal-free region of PI map). 
	Red contour corresponds to a POL-2 total intensity of 13 mJy beam$^{-1}$. The $PI$ is 
	nearly zero in the area surrounding the cores but within the cores themselves a clear detection 
	can be seen. Cyan dashed circles mark the areas with diameters of 
	$3\arcmin$ and $7\arcmin$ around the central positions of the two observed fields. 
	Polarization measurements within the smaller circles as well as the common area covered by both larger circles should 
	be useful. Therefore, the measurements of the three cores T-Tauri, Miz-2, and HGBS-1 may not be reliable due to the dominance of 
	background noise at their locations. Each of the six cores are labelled.}\label{fig:pimap}
\end{figure*}

To extract the reliable data from our POL-2 measurements of the B213 cores, we
adopt the selection criterion $I/\sigma_{I}$~$>$~10 and $P/\sigma_{P}$~$>$~3,
which yield 28 polarization measurements (black circles in Figure \ref{fig:ivspi}).
The resulting median $\sigma_{PI}$ is 0.64 mJy beam$^{-1}$. The $PI$ values lie in the range 1.88 and 3.94 mJy beam$^{-1}$, with a median
of 2.60 mJy beam$^{-1}$, whereas $I$ ranges from 13 to 318 mJy beam$^{-1}$ as shown in
Figure \ref{fig:ivspi}. The $P$ values lie in the range 0.82~--~17.8\% with a minimum $\sim$1\%, median
$\sim$7\% and standard deviation $\sim$5\%. 
Above the 3$\sigma$ level in $PI$, a clear detection of polarized intensity (yellow contours) within the
core boundaries determined from Stokes I map (red contour at 13 mJy beam$^{-1}$)
can be seen in the PI map, as shown in Figure \ref{fig:pimap}. 

\section{B-field at larger scales ($\sim$0.2~--~2.4~pc) determined from optical, near-infrared (NIR), and {\it Planck} polarization data}\label{subsec:largescaleBfields}

In order to compare core-scale B-field (c.f. Section \ref{subsec:smallerBfields})
with that in the large-scale, low-density surrounding region, we make use of archival optical, NIR, and {\it Planck}/850 $\mu$m low-resolution dust polarization data\footnote{{\it Planck} 353 GHz (850$\mu$m) dust continuum polarization data, comprising Stokes $I$, $Q$, and $U$ maps for the B213
region, have been extracted from the {\it Planck} Public Data Release 2 of
Multiple Frequency Cutout Visualization (PR2 Full Mission Map with PCCS2 Catalog: \url{https://irsa.ipac.caltech.edu/applications/planck/)}.
Data have been reduced using the standard procedures described by \citet{PlanckCollaborationXIX2015}, \citet{PlanckCollaborationXXXIII2016a},
\citet{Soleretal2016}, \citet{Baugetal2020} and references therein.};
and the B-field morphologies inferred from these data sets are shown in Figure \ref{fig:b213coldmap}. We select the data within 1$\degr$ of B213; the resulting values of the
Gaussian mean and standard deviation in B-field orientation
($\theta_{B}^{\mathrm{largescale}}\pm \sigma$) are given in Table \ref{tab:cloudpas}.
There exist a significant number of optical/NIR B-field segments around B213; however, NIR polarization measurements are confined to an area to the west of B213 and therefore may not reveal the local B-field of B213. Visual inspection suggests that
optical- and {\it Planck}-inferred B-field is ordered. This is confirmed by their mean B-field orientations which are respectively found to be 29$\pm$14$\degr$ and 29$\pm$17$\degr$. These values are nearly identical, with slightly different standard deviations (cf., Table \ref{tab:cloudpas}),
whereas NIR polarization data show a curved morphology, which follows the
compressed and curved shell of LDN\,1495, with a slightly different mean B-field orientation of 37$\pm$17$\degr$.
Therefore, to delineate the mean B-field in and around B213,
we select the optical and {\it Planck} polarization data within 1$\degr$ of B213, which
yielded a mean orientation of 29$\degr$.
This large-scale, coherent B-field ($\theta_{\mathrm{B}}^{\mathrm{largescale}}$) with a mean orientation of 29$\degr$, 
span spatial scales from $\sim$0.2~pc ($\sim$5$\arcmin$
resolution of {\it Planck}) to $\sim$2.4~pc (1$\degr$ area around B213).


\begin{table*}
        \centering
        \caption{Mean B-field orientations, determined from optical, near-infrared, and low resolution sub-mm 
	({\it Planck}/850\,$\mu$m) polarization observations.}
   \label{tab:cloudpas}
		 \begin{tabular}{ccccc} 
                \hline
\hline
Wavelength &  diameter (arcmin) &   No of stars/segments  & $\theta_{B}^{\mathrm{largescale}}\pm\sigma$ ($\degr$) &  Offset PA$\ddagger$ ($\degr$) \\
(1)        &         (2)        &      (3)               &          (4)                       &  (5)       \\
                                \hline
                Optical$^{a}$                   &  60  &  15 & 29$\pm$14 &  104 \\
                NIR                             &  60  &  42 & 37$\pm$17 &   96 \\
                Sub-mm ({\it Planck}$^{b}$)     &  60  & 445 & 29$\pm$17 &  104 \\
		\hline
\hline
 \end{tabular} \\
 \centering
        \tablecomments{$a$: Two measurements with significant deviation in either $P$ or $\theta$ are excluded from optical data.}
	\tablecomments{$b$: Pixels with values  $<$ 0.008 $K_{\mathrm{CMB}}$  have been excluded from the {\it Planck} data, in order to prevent randomisation of our inferred B-field direction by measurements dominated by noise.} 
\tablecomments{$\theta_{B}^{\mathrm{largescale}}\pm \sigma$ are the mean and standard deviation values resulting from a Gaussian distribution fitted to the data.}
	\tablecomments{Offset PA is difference in angle between the PA of B213 filament ($\sim$133$\degr$) and the large scale 
	mean B-field ($\theta_{B}^{\mathrm{largescale}}$) (column 5).}
\end{table*}

\section{Geometries, effective radii, masses, and column and number densities of B213 cores}\label{appendixsec:geomass}

To estimate various energy and pressure terms for the cores, we extract their masses, column, and number densities from the POL-2 Stokes I map. For this, core dimensions are obtained by fitting the ellipse function {\it mpfitellipse.pro} from the Marquardt library
to the 10$\sigma$ Stokes I contours (13 mJy beam$^{-1}$) of each core.  The resulting core dimensions
(a~$=$~semi-major and b~$=$~semi-minor), effective radius ($R_{\mathrm{eff}}$~$=$~$\sqrt{ab}$), and position angle in degrees east of north are given in Table \ref{tab:paramscl12}.

The integrated fluxes ($F_{\mathrm{\nu}}$)
and median dust temperatures ($T_{\mathrm{d}}$; from the HGBS temperature map) over the core are
used to estimate core masses using the relation \citep{Hildebrand1983}:
\begin {equation}\label{appeq:c1}
M = \frac{F_{\nu}\,D^2}{B_{\mathrm{\nu}}(T_{\mathrm{d}})\,\kappa_\nu}, 
\end{equation}
where $D$~=~140~pc, is the distance of B213, $\kappa_\nu$~$=$~0.0125~cm$^{2}$ g$^{-1}$
\citep[e.g.,][]{Johnstoneetal2017} is the dust mass opacity,
and $B_{\mathrm{\nu}}(T_{\mathrm{d}})$ is Planck function for a blackbody at temperature $T_{\mathrm{d}}$.
The uncertainty in mass is estimated by propagating the standard deviation in $T_{\mathrm{d}}$, 10\% of the value of $F_{\mathrm{\nu}}$ as the flux calibration uncertainty of SCUBA2 \citep{Dempseyetal2013}, 
and a 50\% uncertainty in dust mass opacity \citep[eg.,][]{Royetal2014}.

The column and number densities of the cores are estimated using the following relations:
\begin{equation}\label{eq:coldvold1}
        N(\mathrm{H_{2}}) = \frac{M}{\mu\,m_{\mathrm{H}}\,\pi\,R_{\mathrm{eff}}^2},  \\
\end{equation}
and
\begin{equation}\label{eq:coldvold2}
        n(\mathrm{H_{2}}) = \frac{3\,M}{4\,\mu\,m_{\mathrm{H}}\,\pi\,R_{\mathrm{eff}}^3}. 
\end{equation}

Estimated masses and
column and number densities and their corresponding uncertainties are given along with $T_{\mathrm{d}}$ and $F_{\mathrm{\nu}}$ values in Table \ref{tab:paramscl12}.

\section{Mass-to-flux ratio criticality}\label{subsec:magnetic_criticality}

To infer the importance of B-field with respect to the gravity, we estimate
the mass-to-magnetic flux ratio in units of the critical value
(hereafter mass-to-flux ratio criticality) using the
following relation \citep{Crutcheretal2004,Chapmanetal2011},
\begin{equation}\label{eq:mtf}
        \lambda = \frac{(M/\phi)}{(M/\phi)_{\mathrm{crit}}} =  7.6~N_{\mathrm{\parallel}}(\mathrm{H_{2}})/B_{\mathrm{tot}},
\end{equation}
where $N_{\mathrm{\parallel}}(\mathrm{H_{2}})$ is the mean column density (N(H$_{2}$)$_{\mathrm{POL2}}$),
in units of 10$^{21}$ cm$^{-2}$, along the magnetic flux tube and
$B_{\mathrm{tot}}$ is the total B-field strength in $\mu$G. The critical mass-to-flux ratio,
($M/\phi$)$_{\mathrm{crit}}$~$=$~1/$\sqrt{(4\pi^{2}G)}$ \citep{NakanoNakamura1978}, 
corresponds to the stability criterion for an 
isothermal gaseous layer threaded by perpendicular B-field. A cloud region with
$(M/\phi)$~$>$~$(M/\phi_{\mathrm{crit}})$, i.e. $\lambda$~$>$~1, will collapse under its 
own gravity, and so such a cloud is considered to be supercritical.
A cloud with $\mu$~$<$~1 will be in a subcritical state because of the significant support rendered by B-field.
Taking the mean $N(\mathrm{H_{2}})$~$=$~$N_{\mathrm{\parallel}}(\mathrm{H_{2}})$
as (10$\pm$6)$\times$10$^{21}$, (11$\pm$6)$\times$10$^{21}$~cm$^{-2}$, and (5$\pm$3)$\times$10$^{21}$~cm$^{-2}$
and $B$~$=~B_{\mathrm{tot}}$ a s 38$\pm$17, 48$\pm$22, and 12$\pm$5 $\mu$G, we estimate $\mu$ values of 2$\pm$1, 2$\pm$1, and 3$\pm$2  for K04166, K04169, and Miz-8b, respectively.

However, considering (a) the projection effects between $N_{\mathrm{\parallel}}$(H$_{\mathrm{2}}$)/$B_{\mathrm{tot}}$ and the actual measured N($\mathrm{H_{2}}$)/$B_{\mathrm{\parallel}}$
(where $B_{\parallel}$ is the plane-of-the-sky B-field strength), (ii) B-field being perpendicular to the core elongation in the case of oblate spheroid, or parallel to the core elongation in the case of a prolate spheroid, and
and (iii) the assumption that B-field is randomly oriented with respect to the line of sight, the actual value of $\mu$ becomes (1/3)$\lambda_{\mathrm{obs}}$ for K04166 as the mean B-field is perpendicular to the core major axis; and (3/4)$\lambda_{\mathrm{obs}}$ for K04169 as mean B-field is parallel to the major axis
\citep[][see their Appendix D.4\footnote{In order to correct the estimated mass-to-flux ratio (in critical units) for projection effects, a factor 1/2 is valid for a spheroid cloud, 1/3 for an oblate spheroid flattened
perpendicular to the B-field, and 3/4 for a prolate spheroid elongated along the B-field }]{PlanckCollaborationetal2016}. No correction was applied on the $\lambda$ value of Miz-8b because of the misalignment between mean B-field and core major axis.
Therefore, the resulting $\lambda$ values are 0.7$\pm$0.5, 1.4$\pm$1.0, and 3$\pm$1, which are given in Table \ref{tab:paramscl12}.

\section{Ratio of magnetic-to-rotational energy}\label{appendix:mag2rot}

By assuming that the cores are uniform density spheres and we measure the rotational and magnetic energies using the following relations 
(see \citealt{WursterLewis2020}):
\begin{equation}\label{eq:rotener}
        E_{\mathrm{rot}} = \frac{pMR^{2}\Omega^{2}}{5}, 
        \end{equation}
       and
        \begin{equation}
        E_{\mathrm{mag}} = \frac{B^{2}V}{8\pi},
        \end{equation}
In Equation \ref{eq:rotener},  the correction factor, $p$~$=$~ $\frac{2(3-\mathcal{A})}{3(5-\mathcal{A})}$~$=$~0.27 (where $\mathcal{A}$ is the power-index in the density distribution of the form $\rho \propto r^{-\mathcal{A}}$, here we we consider $\mathcal{A}$~$=$~~1.6), accounts 
for the density distribution in the sphere (see \citealt{XuXetal2020} for more details). 
We use the effective radii $R$~$=$~$R_{\mathrm{eff}}$~$=$~$\sqrt{ab}$ (where $a$~$=$~semi-major axis and $b$~$=$~semi-minor axis), and the volume of the 
core $V$~$=$~(4/3)$\pi R_{\mathrm{eff}}^{3}$. $\Omega$ is the angular velocity or magnitude of
velocity gradient of the core, measured from the N$_{2}$H$^{+}$ data, and is found to be 2.05$\pm$0.02 km s$^{-1}$ pc$^{-1}$ for K04166, 3.86$\pm$0.04 km s$^{-1}$ pc$^{-1}$ for K04169, and 1.88$\pm$0.02 km s$^{-1}$ pc$^{-1}$ for Miz-8b
\citep[][see their Table B.2]{Punanovaetal2018}. $M$ is the mass of the cores (cf., Appendix \ref{appendixsec:geomass}).The derived energy values and their ratio are given in Table \ref{tab:paramscl12}.

\section{Morphological correlation between B-field and the gradients of velocity}\label{appendixsec:bigvg}

We model the observed line-of-sight centroid velocities of 
N$_{2}$H$^{+}$ ($V_{\mathrm{LSR}}$, km s$^{-1}$) and the 
corresponding offset length scales in sky coordinates (R.A ($\Delta_{\alpha}$ in pc) and 
Dec ($\Delta_{\delta}$ in pc)) around each pixel in terms of 
velocity gradients in R.A ($\nabla_{V_{\alpha}}$, km s$^{-1}$ pc$^{-1}$) 
and Dec ($\nabla_{V_{\delta}}$, km s$^{-1}$ pc$^{-1}$) and constant systematic 
velocity of that reference pixel ($V_{0}$, km s$^{-1}$) using the a first-degree bivariate 
polynomial of the form \citep{Goodmanetal1993,Henshawetal2016,Sokolovetal2018} 
\begin{equation}
 V_{\mathrm{LSR}} = V_{\mathrm{0}} + \nabla_{V_{\alpha}} \Delta_{\alpha} + \nabla_{V_{\delta}} \Delta_{\delta}.
 \end{equation} 
We have used the IDL algorithm {\it mpfit} to perform  weighted, non-linear, minimum chi-square 
fitting to constrain the velocity gradients $\nabla_{V_{\alpha}}$ and $\nabla_{V_{\delta}}$, and 
their corresponding uncertainties. 
These are further used to derive the magnitude (${\mathcal{G}}$) and direction ($\Theta_{\mathcal{G}}$) 
of the velocity gradients using the following relations
\begin{equation}
{\mathcal{G}} \equiv |\nabla_{V}| = \sqrt{{\nabla_{V_{\alpha}}}^2 + {\nabla_{V_{\delta}}}^2}
\end{equation}
and
\begin{equation}
\Theta_{\mathcal{G}} = \arctan\left(\frac{{\nabla_{V_{\alpha}}}}{\nabla_{V_{\delta}}}\right).
\end{equation}

We considered at least six adjacent pixels, lying within the beam size of IRAM 30-m telescope for N$_{2}$H$^{+}$ (1~--~0), 26$\farcs$5, around each pixel when fitting was performed. In addition, we estimate the uncertainties in 
the VGs using equation 2 of \citet{Punanovaetal2018}. These were used as weights while performing the weighted fits. The top panels of Figure \ref{fig:igbfield4446} show VGs superimposed on the POL-2 Stokes I maps of K04166, K04169, and Miz-8b. 

\end{document}